\documentclass[a4paper,preprint,aps,prb]{revtex4}
\usepackage{subfigure}
\usepackage{amsmath}
\usepackage{graphicx}
\usepackage{dcolumn}
\usepackage{bm}
\usepackage{color}
\usepackage{lscape}
\usepackage[pdfborder={0 0 0 0}]{hyperref}

\newcommand{\etal}{\it et al.\/ \rm}

\newcommand{\eqa}{\begin{equation}}
\newcommand{\eqz}{\end{equation}}
\newcommand{\eqma}{\begin{eqnarray}}
\newcommand{\eqmz}{\end{eqnarray}}

\begin{document}
\newcommand{\e}{{\em e}~}

\title{Performance of W4 theory for spectroscopic constants and electrical properties of small molecules} 
\author{Amir Karton and Jan M. L. Martin*}
\affiliation{Department of Organic Chemistry, Weizmann Institute of Science, IL-76100 Re\d{h}ovot, Israel\footnote{Present address: Department of Chemistry, University of North Texas, Denton, TX 76203-5017. Email: gershom@unt.edu}}
\email{gershom@weizmann.ac.il}
\date{JCP MS\# A10.07.0057R: submitted July 4, 2010; accepted August 24, 2010}
\begin{abstract}
Accurate spectroscopic constants and electrical properties of small molecules are determined by means of W4 and post-W4 theories. For a set of 28 first- and second-row diatomic molecules for which very accurate experimental spectroscopic constants are available, W4 theory affords near-spectroscopic or better predictions. Specifically, the root-mean-square deviations (RMSD) from experiment are 0.04 pm for the equilibrium bond distances ($r_e$), 1.03 cm$^{-1}$ for the harmonic frequencies ($\omega_e$), 0.20 cm$^{-1}$ for the first anharmonicity constants ($\omega_e x_e$), 0.10 cm$^{-1}$ for the second anharmonicity constants ($\omega_e y_e$), and 0.001 cm$^{-1}$ for the vibration-rotation coupling constants ($\alpha_e$). These RMSD imply 95\% confidence intervals of about 0.1 pm for $r_e$, 2.0 cm$^{-1}$ for $\omega_e$, 0.4 cm$^{-1}$ for $\omega_e x_e$, and 0.2 cm$^{-1}$ for $\omega_e y_e$.
We find that post-CCSD(T) contributions are essential to achieve such narrow confidence intervals for $r_e$ and $\omega_e$, but have little effect on $\omega_e x_e$ and $\alpha_e$, and virtually none on $\omega_e y_e$.
Higher-order connected triples, $\hat{T}_3-$(T), improve agreement with experiment for the hydride systems, but their inclusion (in the absence of $\hat{T}_4$) tends to worsen agreement with experiment for the nonhydride systems. Connected quadruple excitations, $\hat{T}_4$, have significant and systematic effects on $r_e$, $\omega_e$, and $\omega_e x_e$, in particular they universally increase $r_e$ (by up to 0.5 pm), universally reduce $\omega_e$ (by up to 32 cm$^{-1}$), and universally increase $\omega_e x_e$ (by up to 1 cm$^{-1}$). Connected quintuple excitations, $\hat{T}_5$, are spectroscopically significant for $\omega_e$ of the nonhydride systems, affecting $\omega_e$ by up to 4 cm$^{-1}$. Diagonal Born--Oppenheimer corrections have systematic and spectroscopically-significant effects on $r_e$ and $\omega_e$ of the hydride systems, universally increasing $r_e$ by 0.01--0.06 pm and decreasing $\omega_e$ by 0.3--2.1 cm$^{-1}$. Obtaining $r_e$ and $\omega_e$ of the pathologically multireference BN and BeO systems with near-spectroscopic accuracy requires large basis sets in the core-valence CCSD(T) step and augmented basis sets in the valence post-CCSD(T) steps in W4 theory. The triatomic molecules H$_2$O, CO$_2$, and O$_3$ are also considered. The equilibrium geometries and harmonic frequencies (with the exception of the asymmetric stretch of O$_3$) are obtained with near-spectroscopic accuracy at the W4 level. The asymmetric stretch of ozone represents a severe challenge to W4 theory, in particular the connected quadruple contribution converges very slowly with the basis set size. Finally, the importance of post-CCSD(T) correlation effects for electrical properties, namely dipole moments ($\mu$), polarizabilities ($\alpha$), and first hyperpolarizabilities ($\beta$) is evaluated. 
\end{abstract}
\maketitle

\section{Introduction\label{sec:intro}}
Since the early 1990s, there has been growing interest in calculating accurate {\em ab initio} anharmonic force fields for small molecules. There are two aspects to be considered: (i) the nuclear vibrational analysis, and (ii) the level of the electronic structure calculations. For diatomic molecules (i) is relatively straightforward using a Dunham analysis, and for rigid triatomic molecules, such as those considered in the present work, (i) entails no serious problems. As for (ii), for systems with mild to moderate nondynamical correlation effects, the accuracy of CCSD(T) basis set limit results is on the order of 5--10 cm$^{-1}$ for harmonic frequencies\cite{op69,op117, Klo03,Rud04,Tew07} and 0.1--0.5 pm for bond lengths.\cite{op69,op117,Hec05,Hec06} To surpass this level of accuracy (or when considering molecules with more pronounced multireference character such as C$_2$, BN, BeO, and O$_3$) one should consider multireference methods\cite{op34,Set08,Lee08} or post-CCSD(T) contributions.\cite{Rud04,Hec05,Hec06,Tew07,Gau07} Recently, Mintz \etal\cite{Min09} have applied the multireference ccCA method\cite{ccCA} to compute the potential energy curves of N$_2$ and C$_2$; they obtained errors of $\sim$0.06 pm for the bond lengths and $\sim$2 cm$^{-1}$ for the harmonic frequencies. 

The determination of very precise molecular energies (atomization energies or enthalpies of formation) has been one of the primary goals of composite {\em ab initio} methods. It has been shown that highly accurate thermochemical protocols such as HEAT\cite{HEAT1,HEAT2,HEAT3} and W4\cite{op200,op205} are capable of sub-kJ/mol accuracy on average (see Ref.\cite{Tew08} for a recent review). In particular, W4 theory obtains a root mean square deviation (RMSD) of 0.08 kcal/mol against a test set of 25 first- and second-row small molecules for which very accurate ATcT\cite{ATcT} atomization energies are available. This implies a 95\% confidence interval of $\sim$0.16 kcal/mol; the mean signed deviation (MSD) of just -0.01 kcal/mol suggests that W4 is free of systematic bias. Worst-case errors for problematic molecules are $<$ 1 kJ/mol: for example, for F$_2$O$_2$,\cite{op217} C$_2$,\cite{op205} F$_2$O,\cite{op217} and O$_3$,\cite{op200} W4 is 0.09, 0.17, 0.18, and 0.23 kcal/mol away from the ATcT atomization energies (the ATcT value for C$_2$ is available in Ref.\cite{C2-ATcT} and for the remaining species in Ref.\cite{KBHT09}). In our continued work in this field we have shown that contributions from successively higher cluster expansion terms converge increasingly faster with the basis set, as they increasingly reflect nondynamical rather than dynamical correlation.\cite{op200,op205} Indeed, the fact that W4 theory and related methods can be carried out at all at a realistic cost hinges on this behavior. 

In view of the success of W4 theory for thermochemical properties\cite{op200,op205,op217} it is of interest to establish the reliability of W4 (and related) methods for spectroscopic properties based on energy derivatives with respect to the nuclear coordinates ({\em e.g.} $r_e$, $\omega_e$, and $\omega_ex_e$) and for electrical properties based on energy derivatives with respect to an external static electric field ({\em e.g.} $\mu$, $\alpha$, and $\beta$). 

The present study considers a chemically diverse data set of 31 first- and second-row diatomic molecules as well as the triatomic molecules H$_2$O, CO$_2$, and O$_3$. The chosen set, which includes radicals, polar systems, hydrides, and nonhydrides with single and multiple bonds, evidently spans a wide gamut from systems dominated by dynamical correlation ({e.g.} BH and H$_2$O) to systems with pathological nondynamical correlation ({\em e.g.} C$_2$, BN, BeO, and O$_3$) and all shades in between.

\section{Computational methods\label{sec:comp}}
\subsection{Electronic structure}
All the SCF, CCSD, and CCSD(T) calculations were carried out using MOLPRO 2009.1\cite{molpro} running on the Martin group Linux cluster at the Weizmann Institute. The post-CCSD(T) calculations were carried out using MRCC interfaced to the MOLPRO program suite.\cite{MRCC} The diagonal Born--Oppenheimer correction (DBOC) calculations were carried out using the CFOUR program system.\cite{CFOUR} All basis sets employed belong to the correlation consistent family of Dunning and co-workers.\cite{Dun89,Ken92,Wil01,pwCVnZ,Dyall} The notation aug'-cc-pV($n$+d)Z indicates the combination of regular cc-pV$n$Z on hydrogen and aug-cc-pV($n$+d)Z on other elements. 

The computational protocols of W4, W4.2, and W4.3 theories have been specified and rationalized in great detail elsewhere.\cite{op200,op205} In the present work the steps involved in W4 theory are divided in the following manner:
\begin{itemize}
\item W4[up to CCSD(T)$_{fc,nr}$] represents the clamped-nuclei, non-relativistic CCSD(T) infinite basis set limit energy in the frozen-core approximation (in which the 1$s$ orbitals for first-row atoms and the 1$s$, 2$s$, and 2$p$ orbitals for second-row atoms are constrained to be doubly occupied in all configurations). The {\em fc} and {\em nr} subscripts stand for ``frozen-core'' and ``non-relativistic''. The following extrapolations are used for the HF, CCSD, and (T) contributions: 
\begin{itemize}
\item The ROHF-SCF contribution is extrapolated from the aug'-cc-pV(5+d)Z and aug'-cc-pV(6+d)Z basis sets using the Karton-Martin modification\cite{Kar-Mar} of Jensen's extrapolation formula.\cite{Jen05} 
\item The RCCSD valence correlation energy is calculated from these same basis sets. Following the suggestion of Klopper,\cite{Klo01} it is partitioned into singlet-coupled pair energies, triplet-coupled pair energies, and $\hat{T}_1$ terms. The singlet-coupled and triplet-coupled pair energies are extrapolated using the $A+B/L{^\alpha}$	two-point extrapolation formula (where $L$ is the highest angular momentum present in the basis set) with $\alpha_S$=3 and $\alpha_T$=5, respectively, and the $\hat{T}_1$ term (which exhibits very weak basis set dependence) is set equal to that in the largest basis set. 
\item The (T) valence correlation energy is extrapolated from the aug'-cc-pV(Q+d)Z and aug'-cc-pV(5+d)Z basis sets using the $A+B/L{^3}$ two-point extrapolation formula. For open-shell systems the Werner-Knowles-Hampel\cite{Ham2000} (a.k.a. MOLPRO) definition of the restricted open-shell CCSD(T) energy is employed throughout, rather than the original Watts-Gauss-Bartlett\cite{Wat93} (a.k.a. ACES II/CFour) definition.
\end{itemize}
\item W4[up to CCSD(T)] in addition includes inner-shell correlation and scalar relativistic contributions. The former is extrapolated from RCCSD(T)/aug'-cc-pwCVTZ and RCCSD(T)/aug'-cc-pwCVQZ energies using the $A+B/L{^3}$ two-point extrapolation formula, and the latter (in the second-order Douglas-Kroll-Hess approximation\cite{DKH1,DKH2}) is obtained from the difference between non-relativistic CCSD(T)/aug'-pV(Q+d)Z and relativistic CCSD(T)/aug'-cc-pV(Q+d)Z-DK calculations.\cite{Dyall}

\item W4[up to CCSDT] additionally includes higher-order connected triples, $\hat{T}_3-$(T), valence correlation contribution extrapolated from the cc-pVDZ and cc-pVTZ basis sets using the $A+B/L{^3}$ two-point extrapolation formula. 
\item In W4[up to CCSDTQ] the connected quadruples, $\hat{T}_4$, term is also included. The valence correlation (Q) and $\hat{T}_4-$(Q) contributions are calculated with the cc-pVTZ and cc-pVDZ basis sets, respectively. In Refs.\cite{op200,op205} we found that scaling their sum by 1.10 offers a very reliable (as well as fairly cost-effective) estimate of the basis set limit $\hat{T}_4$ contribution.
\item Adding the connected quintuple, $\hat{T}_5$, valence correlation contribution calculated with the $sp$ part of the cc-pVDZ basis set --- denoted cc-pVDZ({no \em d}) --- results in full W4 theory. The $\hat{T}_5$ contribution converges very rapidly with the basis set as it primarily represents static correlation.\cite{op200,op205}
\end{itemize}
W4 represents an approximation to the relativistic basis-set limit CCSDTQ5 energy. The DBOC contributions, calculated at the ROHF/aug'-cc-pVTZ level of theory, are reported separately in Tables \ref{tab:Re}--\ref{tab:ALPHAe}, and are included in the final error statistics presented in these tables.

For the smaller systems we also consider the post-W4 methods (W4.2 and W4.3) as defined in Ref.\cite{op200} The changes in W4.2 and W4.3 relative to W4 are summarized as follows: 
\begin{itemize}
\item W4.2 theory in addition takes account of the $\hat{T}_3-$(T) correction to the core-valence contribution obtained using the cc-pwCVTZ basis set. 
\item In W4.3 all the valence post-CCSD(T) corrections are additionally upgraded: the $\hat{T}_3-$(T) and (Q) corrections are extrapolated from the cc-pVTZ and cc-pVQZ basis sets, the $\hat{T}_4-$(Q) and $T_5$ corrections are calculated with the cc-pVTZ and cc-pVDZ basis sets, respectively, and the $\hat{T}_6$ correction is calculated with the cc-pVDZ({no \em d}) basis set. 
\end{itemize}
Finally, for the pathologically multireference systems in Table \ref{tab:12el} (C$_2$, BN, and BeO) we also consider two additional extensions of W4 theory: 
\begin{itemize}
\item Replacing the regular correlation-consistent basis sets on electronegative atoms (N and O) with their augmented versions in all the valence post-CCSD(T) steps in W4 theory. 
\item Extrapolating the CCSD(T) inner-shell correlation contribution from larger basis sets, namely the aug-cc-pwCVQZ and aug-cc-pwCV5Z basis sets. 
\end{itemize}

\subsection{Vibrational analysis}
For each of the diatomic molecules, a 21-point potential energy curve was calculated using the above-mentioned W4 (and related) theories. The single point energy calculations are carried out at bond distances $r_x=r_e^{expt.}+x$ ($x=0,\pm1, \pm2, \ldots, \pm10$ pm), where $r_e^{expt.}$ is the experimental equilibrium bond length. The spectroscopic constants ($r_e$, $\omega_e$, $\omega_e x_e$, $\omega_e y_e$, and $\alpha_e$) are obtained from a 6$^{th}$--8$^{th}$ degree Dunham fit of the potential energy curves.\cite{Dunham}

For the triatomics CO$_2$, H$_2$O, and O$_3$ grids of all points required to generate the nonvanishing quadratic, cubic, and quartic force constants in symmetry-adapted internal coordinates are calculated using the above-mentioned W4 (and related) theories. 

The convergence criteria throughout are tightened such that the SCF and coupled-cluster energies are converged to at least 10$^{-12}$ a.u.

\subsection{Response properties}
The dipole moments ($\mu$), static polarizabilities ($\alpha$), and static hyperpolarizabilities ($\beta$) are calculated at the experimental structures via numerical differentiation of energies with respect to a uniform electric field. A field strength of $h$=0.0025 a.u. was chosen as the basic grid size: the values reported were obtained by means of
Richardson extrapolation\cite{RalRab} from successive multiples of that grid size.
 The first-order ($\mu_z$) and third-order ($\beta_{zzz}$) partial derivatives are obtained by means of a four point formula using static field strengths of $\pm h$ and $\pm 2h$. The second-order derivatives ($\alpha_{xx}$, $\alpha_{yy}$, and $\alpha_{zz}$) are obtained through a five point formula using field strengths of 0, $\pm h$, and $\pm 2h$. The mixed third-order derivatives ($\beta_{jjz}, ~j = x, y$) are computed using a six point formula with field strengths of $\pm 2h{\bf \hat{z}}$, $\pm 2h( {\bf\hat{z}}+{\bf\hat{j}})$, and $\pm 2h( {\bf\hat{z}}- {\bf\hat{j}})$.

\section{Results and discussion\label{sec:results}}
\subsection{Multireference considerations\label{subsec:ndc}}
In reference\cite{op200} we proposed a number of energy-based diagnostics for the importance of nondynamical correlation that are specifically designed for thermochemical purposes. In particular, the \%TAE$_e$[(T)] diagnostic, {\em i.e.} the percentage of the total atomization energy accounted for by parenthetical connected triple excitations, proved to be a very efficient and cost-effective {\em a priori} diagnostic:
\begin{equation}
\%{\rm TAE_e}[(T)] = 100\times\frac{{\rm TAE_{\it e}[CCSD(T)]}-{\rm TAE_{\it e}[CCSD]}}{{\rm TAE_{\it e}[CCSD(T)]}}
\end{equation}
where TAE$_e$[CCSD] and TAE$_e$[CCSD(T)] represent, respectively, the non-relativistic, clamped-nuclei CCSD and CCSD(T) total atomization energies at the bottom of the well. Values of \%TAE$_e$[(T)] are interpreted as follows:\cite{op200} below 2\% indicates systems dominated by dynamical correlation; 2--5\% mild nondynamical correlation; 5--10\% moderate nondynamical correlation; and in excess of 10\% severe nondynamical correlation. 

Table S-I of the Supporting Information\cite{SuppInfo} provides the percentage of the non-relativistic, clamped-nuclei total atomization energy at the bottom of the well accounted for by SCF, (T)	triples, post-CCSD(T), and	$\hat{T}_4+\hat{T}_5$ contributions. Table S-I also lists the coupled cluster ${\cal T}_1$ diagnostic,\cite{T1} ${\cal D}_1$ diagnostic,\cite{D1} and the largest CCSD $T_2$ amplitudes. However, the use of these diagnostics as a measure of multireference effects has been criticized by us\cite{op200,op217} and by others.\cite{Wat94,Wat95,Truh07} In the context of the present work, for example, a ${\cal T}_1$ diagnostic of 0.03 is obtained for CS and SiO as well as for O$_3$, although the latter is a notoriously multireference system. The ${\cal D}_1$ diagnostic of O$_2$ (0.01) is the same as, or even smaller than, that of the hydride systems, despite the former having almost an order of magnitude greater \%TAE$_e$[$\hat{T}_4+\hat{T}_5$]. The largest $T_2$ cluster amplitudes are deceptively high for BH (0.11) and deceptively low for BeO (0.04). In practice, the \%TAE[(T)] indicator proves to be a very useful predictor for the importance of post-CCSD(T) correlation effects (see relevant discussions in Refs.\cite{op200,op217}). 
 
As expected, the hydride systems considered in the present study exhibit very mild nondynamical correlation effects and can be regarded as dominated by dynamical correlation: 60--80\% of the atomization energy is accounted for at the SCF level, and only 0--2\% by the (T) triples. The nonhydrides BF, BCl, AlF, AlCl, CO, SiO, SiS, and N$_2$ are characterized by mild nondynamical correlation: at least 50\% of the TAE$_e$ is accounted for at the SCF level, and 2$\leq$\%TAE$_e$[(T)]$\leq$5. The homonuclear diatomics (O$_2$, Si$_2$, P$_2$, S$_2$, and Cl$_2$), as well as CS, PN, SO, ClF, and BeO are characterized by moderate nondynamical correlation effects, 20$\leq$\%TAE$_e$[SCF]$\leq$60 and 6$\leq$\%TAE$_e$[(T)]$\leq$8. Finally, C$_2$, BN, and O$_3$ are pathologically multireference: at the SCF level BN and O$_3$ are unbound, and only 10\% of the atomization energy of C$_2$ is accounted for, while the (T) triples account for as much as 13--19\% of the binding energies of these systems.

\subsection{Equilibrium geometries ($r_e$) of diatomic molecules\label{subsec:Re}}
Table \ref{tab:Re} compares the equilibrium bond distances obtained at the W4 (and related) levels with experimental bond lengths for our set of 10 monohydride and 18 nonhydride diatomic molecules. The equilibrium bond lengths determined at the clamped-nuclei, frozen-core, non-relativistic CCSD(T) basis set limit (W4[up to CCSD(T)$_{fc,nr}$] method) almost universally overestimate the experimental bond distances (by $\sim$0.3 pm on average). Inclusion of the core-valence and scalar relativistic corrections consistently reduces $r_e$ by amounts ranging from $\sim$0.1 pm (for HF, OH, F$_2$, and NH) to $\sim$1.0 pm (for AlCl and Si$_2$). Consequently, at the clamped-nuclei, all-electron, relativistic CCSD(T) basis set limit (W4[up to CCSD(T)] method) $r_e$ are universally underestimated (by $\sim$0.2 pm on average). The largest underestimations are seen for the nonhydrides, {\em e.g.} by 0.26, 0.28, 0.31, 0.32, 0.33, 0.35, 0.38, 0.39, and 0.52 pm, for SiS, PN, ClF, Cl$_2$, O$_2$, S$_2$, F$_2$, P$_2$, and Si$_2$, respectively. For the subset of 18 nonhydride systems the W4[up to CCSD(T)] method attains an RMSD of 0.27 pm. As expected, the performance for the subset of monohydride systems, which are mostly dominated by dynamical correlation (0.1$\leq$\%TAE[(T)]$\leq$1.6), is much better with an RMSD of only 0.06 pm. For the entire set of 28 diatomics W4[up to CCSD(T)] theory attains an RMSD of 0.22 pm. 

Higher-order triple excitations, $\hat{T}_3-$(T), have a relatively small effect on the bond lengths: $|r_e|$ changes by $\leq$0.1 pm (with the exception of P$_2$ and PN for which $r_e$ decreases by 0.2 pm). For the hydride systems the $\hat{T}_3-$(T) contribution consistently increases the bond lengths by 0.01--0.06 pm thereby improving agreement with experiment. For the nonhydride systems the $\hat{T}_3-$(T) excitations have a mixed effect, bettering agreement with experiment for some while worsening it for others. For the entire set of 28 diatomics, the effect of including the $\hat{T}_3-$(T) contribution is manifested by a slight increase in the RMSD from 0.22 to 0.25 pm. Note that, with very few exceptions (namely BH, CH, and AlH), the W4[up to CCSDT] method still universally underestimates the experimental bond lengths. 

Connected quadruple excitations, $\hat{T}_4$, universally increase the bond distances, thereby substantially improving the agreement with experiment, particularly for the nonhydride systems. Upon inclusion of the $\hat{T}_4$ excitations the RMSD over the nonhydrides is reduced from 0.30 to 0.06 pm (!), where in most cases the bond lengths increase by $>$ 0.2 pm ({\em e.g.} by 0.28, 0.29, 0.30, 0.33, 0.37, 0.42, and 0.46 for SO, S$_2$, Si$_2$, O$_2$, PN, F$_2$, and P$_2$, respectively). For the monohydrides, the effect of the $\hat{T}_4$ excitations is rather modest, the bond lengths increasing by only 0.00--0.03 pm and the RMSD being reduced from 0.04 to 0.03 pm. 

As expected, connected quintuple excitations have essentially no effect on the internuclear distances of the hydride systems. For the nonhydride systems the bond lengths generally increase by 0.00--0.05 pm upon inclusion of the $\hat{T}_5$ excitations. 

At the full W4 level the RMSD from experiment are \{0.03, 0.05, 0.04\} pm for the hydrides, nonhydrides, and the entire set, respectively. The largest deviations (underestimations) are 0.07 pm (for P$_2$, SiH, SiS, and ClF) and 0.09 pm (for Cl$_2$).

The trends discussed above can be qualitatively understood by looking at how the various energy contributions ($E^{{\rm core-valence}}$, $E^{\hat{T}_3-{\rm (T)}}$, $E^{\hat{T}_4}$, and $E^{\hat{T}_5}$) change as a function of $r$. Tables SII--SV of the Supporting Information\cite{SuppInfo} give these energy contributions as a function of $r$. The said energy contributions (with the exception of $E^{\hat{T}_3-{\rm (T)}}$ for BF and AlF; {\em vide infra}) show a nearly linear dependence on $r$ in the interval $|r-r_e|\leq 10$ pm. Addition of a linear term ($E\approx ar$) to a simple harmonic potential results in an increase in $r_e$ if $a<0$ and a decrease in $r_e$ if $a>0$. For all the systems, the connected quadruples energy contribution, $E^{\hat{T}_4}$, decreases with $r$ (in a nearly linear manner: $R^2\geq0.99$) and thereby inclusion of the $\hat{T}_4$ excitations universally lengthens the bond distances. The core-valence energy contribution, $E^{{\rm core-valence}}$, systematically increases with $r$ (in a fairly linear fashion, $R^2\geq0.98$), thus universally shortening the bond lengths. For the hydride systems $E^{\hat{T}_3-{\rm (T)}}$ decreases linearly with $r$ ($R^2\geq0.98$), therefore lengthening the equilibrium bond lengths. For the nonhydrides, the $E^{\hat{T}_3-{\rm (T)}}$ energy contribution does not behave in a consistent manner: for some (namely Si$_2$, Cl$_2$, BCl, ClF, AlCl, and SiS) it decreases with $r$ therefore increasing $r_e$, while for others it increases with $r$, thus decreasing $r_e$. Interestingly, for the highly polar BF and AlF systems $E^{\hat{T}_3-{\rm (T)}}$ varies quadratically with $r$ in the scanned interval ($R^2\geq0.998$, see Table S-II of the Supporting Information\cite{SuppInfo}) and $r_e$ is not affected by the $\hat{T}_3-$(T) correction. For most of the systems $E^{\hat{T}_5}$ decreases with $r$ (exceptions are: BF, BCl, AlF, AlCl, CS, SiS, and SiO), therefore inclusion of $\hat{T}_5$ excitations increases $r_e$ for all but the said exceptions. In the next section we shall see that the said energy contributions that increase $r_e$ decrease $\omega_e$ and vice versa --- as expected since $\omega\propto\frac{1}{r}$. 

For a few systems we were able to obtain potential curves at the W4.2 and W4.3 levels (see Table \ref{tab:Re}). Adding the $\hat{T}_3-$(T) correction to the core-valence contribution in W4.2 theory has little effect on the bond distances, namely $r_e$ is increased by 0.02 and 0.03 pm for F$_2$ and P$_2$, respectively. Likewise, computing all the valence post-CCSD(T) corrections with larger basis sets and adding a $\hat{T}_6$ correction in W4.3 theory increases the bond length of F$_2$ by merely 0.03 pm. 

Finally, DBOC corrections universally increase the bond distances and, as expected, only have a perceptible effect on the hydride systems, something more pronounced for the lighter hydrides within each row. For example, the DBOC correction increases $r_e$ by 0.06, 0.04, 0.03, 0.02, and 0.01 pm for the first-row hydrides BH, CH, NH, OH, and HF, respectively, and by 0.05, 0.04, 0.03, 0.02, and 0.01 pm for the second-row hydrides AlH, SiH, PH, SH, and HCl, respectively.

\subsection{Harmonic frequencies ($\omega_e$) of diatomic molecules\label{subsec:We}}
The harmonic frequencies at the W4 (and related) levels are given in Table \ref{tab:We} together with the experimental values. At the clamped-nuclei, all-electron, relativistic CCSD(T) basis set limit (W4[up to CCSD(T)] method) the harmonic frequencies are universally overestimated by amounts ranging from $\sim$1 cm$^{-1}$ (for AlF and AlCl) up to 29 cm$^{-1}$ (for O$_2$). Not surprisingly, the smallest errors (of 1--8 cm$^{-1}$) are observed for the systems for which \%TAE$_e$[(T)]$<$4 (see section \ref{subsec:ndc}). The RMSD of the W4[up to CCSD(T)] method are \{5.1, 11.2, 9.8\} cm$^{-1}$ for the hydrides, nonhydrides, and the entire set, respectively.

For most systems W4[up to CCSD(T)$_{fc,nr}$] and W4[up to CCSD(T)] err on the same side of experiment, the exceptions are AlF, CO, BCl, CH, BF, and BH for which W4[up to CCSD(T)$_{fc,nr}$] underestimates experiment by 2.5--8.6 cm$^{-1}$. We note that with few exceptions (namely BH, AlH, BF, AlF, CH, and SiH) W4[up to CCSD(T)$_{fc,nr}$] is closer to experiment than W4[up to CCSD(T)], as manifested in an RMSD for the entire set smaller by 2 cm$^{-1}$. 

Higher-order triple excitations, $\hat{T}_3-$(T), consistently reduce the harmonic frequencies of the monohydride compounds by 1.4--5.0 cm$^{-1}$, thus cutting the RMSD for the monohydrides by almost 50\% from 5.1 cm$^{-1}$ (W4[up to CCSD(T)]) to 2.8 cm$^{-1}$ (W4[up to CCSDT]). For the nonhydrides, the higher-order triple excitations have a mixed effect, reducing $\omega_e$ for some systems (e.g., by $\sim$1 cm$^{-1}$ for Si$_2$, BCl, and Cl$_2$) and increasing $\omega_e$ for others (most notably by $\sim$10 cm$^{-1}$ for N$_2$ and PN). In effect, the RMSD for the nonhydrides increases upon inclusion of the $\hat{T}_3-$(T) correction (from 11.2 cm$^{-1}$ for W4[up to CCSD(T)] to 15.2 cm$^{-1}$ for W4[up to CCSDT]). 

Connected quadruple excitations, $\hat{T}_4$, universally reduce $\omega_e$ by amounts ranging from 0.0--5.0 cm$^{-1}$ for the monohydride systems and 0.3--31.5 cm$^{-1}$ for the nonhydride systems. The largest decreases, of over 10 cm$^{-1}$, are seen for P$_2$, SO, F$_2$, N$_2$, PN, and O$_2$ (specifically by 13.4, 16.1, 16.7, 20.4, 21.8, and 31.5 cm$^{-1}$, respectively). Inclusion of the $\hat{T}_4$ excitations reduces the RMSD over the set of hydrides from 2.8 cm$^{-1}$ (W4[up to CCSDT]) to 0.6 cm$^{-1}$ (W4[up to CCSDTQ]), and over the set of nonhydrides from 15.2 cm$^{-1}$ (W4[up to CCSDT]) to 2.4 cm$^{-1}$ (W4[up to CCSDTQ]). Over the entire set of 28 diatomics the RMSD is reduced from 12.7 cm$^{-1}$ (W4[up to CCSDT]) to 2.0 cm$^{-1}$ (W4[up to CCSDTQ]), an improvement by nearly one order of magnitude (!). This clearly demonstrates that connected quadruple excitations are essential for obtaining the harmonic frequencies with near-spectroscopic accuracy. 

As expected, connected quintuple excitations, $\hat{T}_5$, have little or no effect for the hydrides: the biggest changes are seen for OH and HF for which $\omega_e$ is reduced by merely 0.1 cm$^{-1}$. For the nonhydrides $\hat{T}_5$ excitations change $\omega_e$ by up to 4 cm$^{-1}$. The biggest changes are observed for N$_2$, PN, and O$_2$, namely, $\omega_e$ is reduced by 3.3, 3.5, and 4.0 cm$^{-1}$, respectively. 

Upon inclusion of the $\hat{T}_5$ excitations the RMSDs for the nonhydrides (and for the entire set) are reduced by 50\%, thus at the W4 level RMSDs of \{0.6, 1.2, 1.0\} cm$^{-1}$ are obtained for the hydrides, nonhydrides, and the entire set, respectively. The largest deviation being 2.1 cm$^{-1}$ for F$_2$ and P$_2$. We note that these errors drop to 0.9 and 1.3 cm$^{-1}$, respectively, upon adding the $\hat{T}_3-$(T) correction to the core-valence contribution in W4.2 theory. The error for F$_2$ is further reduced to 0.5 cm$^{-1}$ by upgrading the valence post-CCSD(T) corrections and adding a $\hat{T}_6$ correction in W4.3 theory. 

Finally, a word is due on the effect of the DBOC correction on the harmonic frequencies. For the monohydride systems the DBOC reduces $\omega_e$ across the board, again more appreciably so for the lighter hydrides within each row. Specifically, the DBOC reduces $\omega_e$ by 2.1, 1.8, 1.4, 0.9, and 0.3 cm$^{-1}$ for the first-row hydrides BH, CH, NH, OH, and HF, respectively, and by 1.1, 1.0, 0.9, 0.7, and 0.5 cm$^{-1}$ for the second-row hydrides AlH, SiH, PH, SH, and HCl, respectively. For the nonhydrides the DBOC negligibly increases the harmonic frequencies by $<$0.01 cm$^{-1}$, the biggest change of +0.08 cm$^{-1}$ being found for O$_2$.

\subsection{First anharmonicity constant ($\omega_e x_e$) of diatomic molecules\label{subsec:WeXe}}
Table \ref{tab:WeXe} lists the theoretical and experimental first anharmonicity constants. Monohydrides are well known to exhibit relatively strong anharmonicity effects: $\omega_e x_e$ ranges between 30--90 cm$^{-1}$ for the subset of monohydrides, and between 2--14 cm$^{-1}$ for the subset of nonhydrides.

The W4[up to CCSD(T)] method almost universally underestimates $\omega_e x_e$ by amounts ranging from 0.01 cm$^{-1}$ (BF) to 1.0 cm$^{-1}$ (NH). At this level RMSDs of \{0.6, 0.3, 0.5\} cm$^{-1}$ are obtained for the hydrides, nonhydrides, and the entire set, respectively. We note that the W4[up to CCSD(T)$_{fc,nr}$] method shows very similar performance for most of the systems, the core-valence correction tends to slightly increase $\omega_e x_e$ but overall offers little improvement. 

Inclusion of higher-order triple excitations improves agreement with experiment for the monohydride systems, and slightly worsens agreement with experiment for the nonhydride systems. For the monohydride systems $\hat{T}_3-$(T) excitations increase $\omega_e x_e$ by up to 0.5 cm$^{-1}$ (for NH), whereas for the nonhydrides they tend to decrease $\omega_e x_e$ by up to 0.3 cm$^{-1}$ (for N$_2$ and F$_2$). The RMSD of the W4[up to CCSDT] method is 0.4 cm$^{-1}$ for both the hydride and nonhydride subsets. 

Inclusion of connected quadruple excitations in the W4[up to CCSDTQ] method universally increases $\omega_e x_e$ by up to 1.0 cm$^{-1}$. Thus, improving agreement with experiment across the board, something more pronounced for the nonhydride systems. The largest improvements are seen for N$_2$ and O$_2$: the error for N$_2$ is reduced from 0.7 cm$^{-1}$ to 0.2 cm$^{-1}$, and that for O$_2$ from 1.1 cm$^{-1}$ to 0.1 cm$^{-1}$. 

Contributions from connected quintuple excitations are generally negligible (for most of the systems $<$0.01 cm$^{-1}$) the largest changes being +0.06, +0.11, and +0.15 cm$^{-1}$ for PN, N$_2$, and O$_2$, respectively. 

For the nonhydride systems agreement between W4 and experiment is as good as one can hope for with an RMSD of 0.04 cm$^{-1}$. For the hydride systems the RMSD is almost one order of magnitude larger (0.32 cm$^{-1}$) but still well within the goal of ``spectroscopic accuracy''. 

W4.2 and W4.3 theories offer no significant improvement over W4 theory. Likewise, DBOC corrections are negligible (the largest correction of -0.02 cm$^{-1}$ is obtained for BH).

\subsection{Second anharmonicity constant ($\omega_e y_e$) of diatomic molecules\label{subsec:WeYe}}
Table \ref{tab:WeYe} reports the theoretical and experimental second anharmonicity constants. The second-order anharmonicity constants are typically 2--3 orders of magnitude smaller than the first-order anharmonicity constants. For the monohydrides $\omega_e y_e$ ranges from 0.1 cm$^{-1}$ (for SiH, PH, and SH) to 0.9 cm$^{-1}$ (for HF). For the nonhydrides, which exhibit weaker anharmonicity effects, $\omega_e y_e$ are typically $<$0.01 cm$^{-1}$, and are given in Table S-VI of the Supporting Information.

The W4[up to CCSD(T)$_{fc,nr}$] method attains near-zero RMSD of 0.003, 0.09, and 0.06 cm$^{-1}$ for the nonhydrides, hydrides, and the entire set, respectively. For both subsets core-valence, post-CCSD(T), and DBOC corrections have very little effect on the second anharmonicity constants.

\begin{figure}
\includegraphics[scale=1.0]{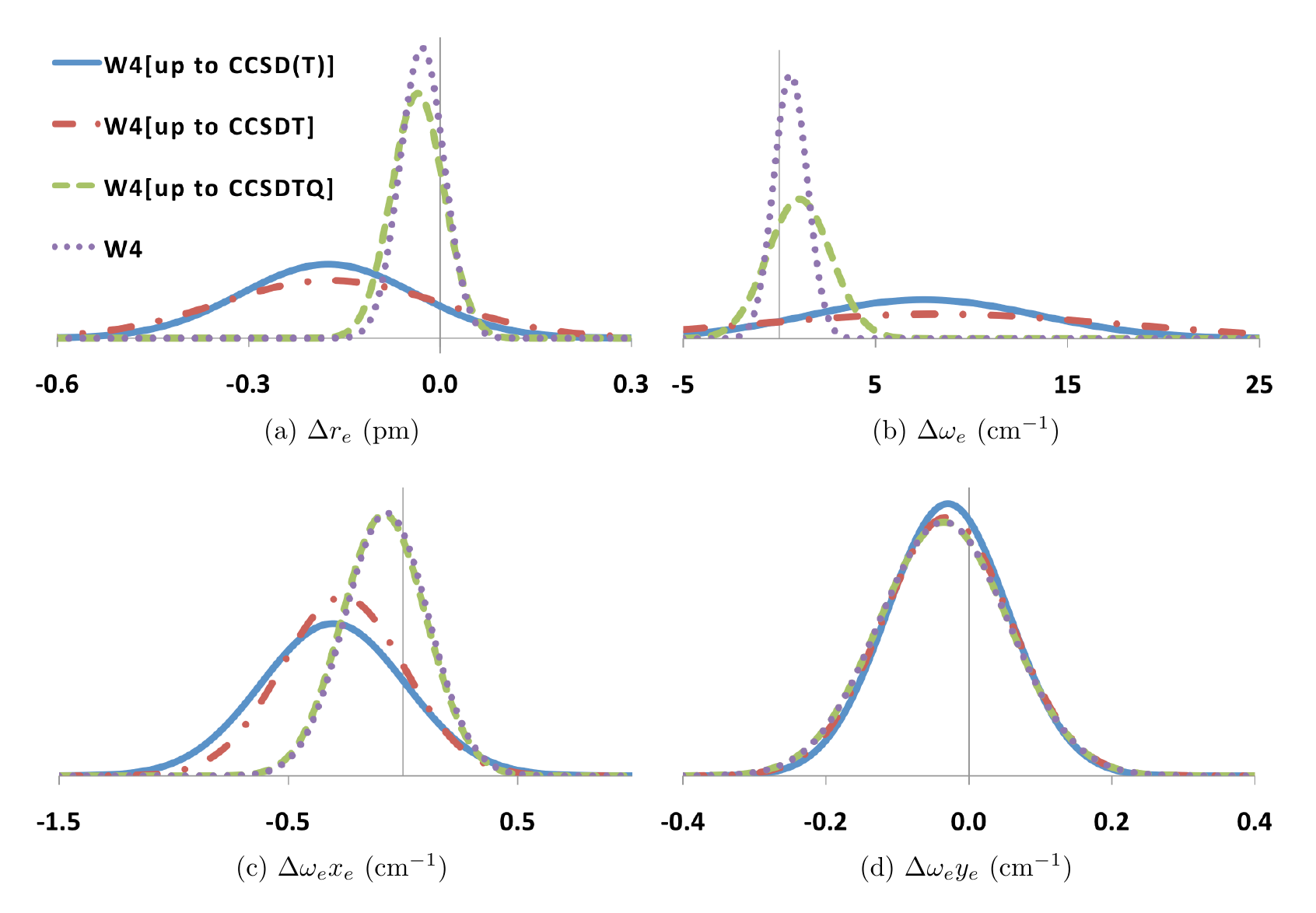}
\caption{Normal distribution of errors between theory and experiment for $r_e$, $\omega_e$, $\omega_e x_e$, and $\omega_e y_e$. The centers of the Gaussians for the W4[up to CCSD(T)], W4[up to CCSDT], W4[up to CCSDTQ], and W4 methods, respectively, are given in parenthesis. (a) equilibrium bond lengths (-0.16, -0.17, -0.04, -0.03 pm), (b) harmonic frequencies (7.5, 8.1, 1.0, and 0.6 cm$^{-1}$), (c) first-order anharmonic corrections (-0.30, -0.26, -0.08, -0.07 cm$^{-1}$), and (d) second-order anharmonic corrections (-0.03, -0.03, -0.04, -0.04 cm$^{-1}$). \label{fig:1}}
\end{figure}

Figure \ref{fig:1} shows the normal distribution functions for the errors between the theoretical and experimental $r_e$, $\omega_e$, $\omega_e x_e$, and $\omega_e y_e$. It illustrates that $r_e$ and $\omega_e$ are much more sensitive to post-CCSD(T) correlation effects, in particular to the $\hat{T}_4$ excitations, than the anharmonic corrections. This was suggested long ago, based on CEPA\cite{cepa} calculations on the methane surface by Pulay and coworkers.\cite{Pulay78} The Gaussians for the W4[up to CCSDT] and W4[up to CCSDTQ] methods are centered, respectively, around -0.17 and -0.04 pm for $r_e$, around 8.1 and 1.0 cm$^{-1}$ for $\omega_e$, around -0.3 and -0.1 cm$^{-1}$ for $\omega_e x_e$, and around -0.03 and -0.04 cm$^{-1}$ for $\omega_e y_e$.

\subsection{Vibration-rotation coupling constant ($\alpha_e$) of diatomic molecules\label{subsec:ALPHAe}}
Table \ref{tab:ALPHAe} depicts the theoretical and experimental vibration-rotation coupling constants. 
The RMSD of the W4[up to CCSD(T)$_{fc,nr}$], W4[up to CCSD(T)], W4[up to CCSDT], W4[up to CCSDTQ], and W4 methods are 0.0034, 0.0024, 0.0017, 0.0012, and 0.0012 cm$^{-1}$, respectively. Hence it can be concluded that, while the $\alpha_e$ are reasonably well converged at the clamped-nuclei, frozen-core, non-relativistic CCSD(T) basis set limit (W4[CCSD(T)$_{fc,nr}$] method), significant improvements can still be obtained beyond it.

\subsection{Difficult 8-valence electron systems: C$_2$, BN, and BeO\label{subsec:12el}}
The low-lying $^1\Sigma^+$ electronic states of the twelve-electron isoelectronic diatomics C$_2$, BN, and BeO are known to have significant contributions from singly and doubly excited determinants, and thus represent an extremely challenging test for single reference electron correlation methods.\cite{Wat94} Table \ref{tab:12el} lists the spectroscopic constants at the W4 (and related) levels together with the experimental values. The harmonic frequencies are severely overestimated at the W4[up to CCSD(T)] level, specifically by 17, 50, and 10 cm$^{-1}$ for C$_2$, BN, and BeO, respectively. Higher-order triples, $\hat{T}_3-$(T), contributions increase the errors for C$_2$ and BeO to about 20 cm$^{-1}$, but reduce the error for BN by 36 cm$^{-1}$ (!) to 14 cm$^{-1}$. (Interestingly, the same decrease of 36 cm$^{-1}$ was computed nearly two decades ago by one of us using a double zeta quality basis set.\cite{op34}) As expected, connected quadruple excitations reduce $\omega_e$ by substantial amounts, namely 15.5, 6.6, and 12.6 cm$^{-1}$ for C$_2$, BN, and BeO, respectively. Note however, that for BN the $\hat{T}_4$ excitations have a lesser effect on $\omega_e$ than the $\hat{T}_3-$(T) excitations. $\hat{T}_5$ excitations further reduce the errors for C$_2$ and BN by 1.5 and 2.0 cm$^{-1}$, respectively, but increase the error for BeO by 7.2 cm$^{-1}$. Thus, at the W4 level the harmonic frequencies of C$_2$, BN, and BeO are overestimated by 2.6, 5.7, and 14.3 cm$^{-1}$, respectively. 

We suspected that the large discrepancies seen for the ionic BN and BeO molecules are partly due to the difficulty of the unaugmented basis sets to describe the charge distribution in the valence post-CCSD(T) steps. Adding diffuse functions on the electronegative atoms (N and O) in all the valence post-CCSD(T) steps in W4 theory (see Computational Methods section and footnote $b$ in Table \ref{tab:12el}) reduces the errors for BN and BeO to 2.1 and 6.1 cm$^{-1}$, respectively. In addition, because of the fairly small core-valence gap in BeO, we also considered extrapolating the CCSD(T) inner-shell correlation contribution from larger basis sets (see Computational Methods section and footnote $c$ in Table \ref{tab:12el}), which results in further reduction of the error for BeO to 1.2 cm$^{-1}$. (The reduction to 1.9 cm$^{-1}$ for BN is not significant.)

As mentioned in section \ref{subsec:Re}, the W4[up to CCSD(T)] method consistently underestimates the experimental bond lengths of the diatomic molecules in Table \ref{tab:Re} (by $\sim$0.2 pm on average). Similarly, the bond distances of C$_2$ and BeO are underestimated by 0.2 pm, but that of BN is underestimated by as much as 1.0 pm. Higher-order triples, $\hat{T}_3-$(T), have little effect on $r_e$ of C$_2$ and BeO, but increase $r_e$ of BN by as much as 0.7 pm (!) narrowing the gap between theory and experiment to 0.3 pm. $\hat{T}_4$ contributions further reduce the discrepancy between theory and experiment: W4[up to CCSDTQ] underestimates the experimental bond lengths of C$_2$, BN, and BeO by 0.05, 0.12, 0.11 pm, respectively. $\hat{T}_5$ excitations increase $r_e$ of C$_2$ and BN by 0.03 and 0.05 pm, but decrease $r_e$ of BeO by 0.11 pm. Thus, at the W4 level the C$_2$ and BN bond lengths are in relatively good agreement with experiment (underestimating by 0.03 and 0.07 pm, respectively), but $r_e$ of BeO is underestimated by as much as 0.21 pm. Again, augmenting the basis sets on N and O in all the valence post-CCSD(T) steps in W4 theory reduces the discrepancy for BeO to 0.08 pm, and for BN to 0.03 pm. Extrapolating the inner-shell correlation from RCCSD(T)/aug-cc-pwCVQZ and RCCSD(T)/aug-cc-pwCV5Z calculations results in further reduction of the error for BN and BeO to merely 0.02 pm. 

Turning our attention to the first anharmonicity constants, for C$_2$ errors of 1.1 and 0.1 cm$^{-1}$ are obtained at the W4[up to CCSD(T)] and W4 levels, where the biggest improvement comes from the $\hat{T}_4$ excitations (see Table \ref{tab:12el}). For BN errors of 5.5 and 1.2 cm$^{-1}$ are obtained at the W4[up to CCSD(T)] and W4 levels, but here the improvement essentially stems from the $\hat{T}_3-$(T) contribution. For BeO errors of 0.3 and 0.5 cm$^{-1}$ are obtained at the W4[up to CCSD(T)] and W4 levels. We note that by adding diffuse functions on N and O in all the post-CCSD(T) steps in W4 theory the errors for BN and BeO are further reduced to 0.85 and 0.05 cm$^{-1}$, respectively.

\subsection{Triatomic systems}
The W4 equilibrium geometries, harmonic frequencies, and fundamental frequencies of H$_2$O, CO$_2$, and O$_3$ are given in Table \ref{tab:tri} together with the available experimental values. The W4 equilibrium geometries are in close agreement with the experimental ones. The equilibrium bond distances of CO$_2$ and O$_3$ are both 0.02 pm lower than experiment, and the bond angle of O$_3$ is 0.07$^\circ$ higher than that of Barbe {\em et al}.\cite{Bar02} For H$_2$O the W4 bond distance (95.76 pm) is bracketed in between the experimental value of Benedict \etal\cite{Ply56} (95.72 pm) and the more recent value of Jensen\cite{Jen88} (95.84 pm), and the W4 bond angle (104.50$^\circ$) is in between the said experimental values (104.52$^\circ$\cite{Ply56} and 104.44$^\circ$\cite{Jen88}, respectively). The Born--Oppenheimer correction is expected to be negligible for CO$_2$ and O$_3$, and is on the order of 0.003 pm and 0.02$^\circ$ for the bond distance and bond angle of water, respectively.\cite{Csa05} We note that our W4 $r_e$ and $\theta_e$ for water differ by 0.02 pm and 0.02$^\circ$, respectively, from the {\em ab initio} Born--Oppenheimer geometry reported in Ref.\cite{Csa05}.

The W4 harmonic frequencies for CO$_2$ and H$_2$O are in relatively good agreement with experiment. In particular, the symmetric, bending, and antisymmetric vibrational frequencies of CO$_2$ underestimate the experimental values by 1.9, 2.4, and 0.1 cm$^{-1}$, respectively, and those of H$_2$O overestimate experiment\cite{Jen88} by 2.8, 0.5, and 3.3 cm$^{-1}$, respectively. 

Ozone represents an extremely challenging system for both single reference and multireference methods (see, {\em e.g.} \cite{Sab04,Sta03,Sta94} and references therein). The theoretical symmetric stretch converges smoothly with the level of theory, {\em i.e.} it overestimates the experimental value of Barbe \etal\cite{Bar02} by 20.3, 3.7, and 0.6 cm$^{-1}$, respectively, at the W4[up to CCSDT], W4[up to CCSDTQ], and W4 levels. The harmonic bending frequency converges more rapidly, in particular it overestimates the experimental one by 2.9, 0.5, and 0.2 cm$^{-1}$ at the same levels, respectively. However, the asymmetric stretch ($\omega_3$), which exhibits severe nondynamical correlation character\cite{Lee87,Sta89,Bart99}, behaves in a less systematic manner. It underestimates the experimental value by 1.6, 17.1, and 20.2 cm$^{-1}$, respectively, at the said sequence of levels (where the good agreement between the W4[up to CCSDT] value and experiment is clearly fortuitous). It is likely that the counterintuitive deterioration in agreement with experiment upon inclusion of the connected quadruple excitations is due to basis set deficiencies. The hypersensitivity of $\omega_3$ to the treatment of connected quadruple excitations has been noted in Ref.\cite{Bart99} Determination of $\omega_3$ requires evaluation of the energy at only one point (in $C_s$ symmetry) aside from the equilibrium $C_{2v}$ structure. We were able to obtain the CCSDTQ/cc-pVTZ({no \em f 1d}) energies at these two points --- cc-pVTZ({no \em f 1d}) denotes the $sp$ part of the cc-pVTZ basis set combined with the $d$ function from the cc-pVDZ basis set. Replacing the $\hat{T}_4$ term in W4 theory with $\hat{T}_4$/cc-pVTZ({no \em f 1d}) results in an asymmetric stretch of 1078.1 cm$^{-1}$, {\em i.e.} the error from experiment is reduced by more than 50\%. Additionally upgrading the basis set for the parenthetical connected quadruple excitations to cc-pVTZ ({\em i.e.}, $\hat{T}_4$=CCSDT(Q)/cc-pVTZ--CCSDT/cc-pVTZ+CCSDTQ/cc-pVTZ({no \em f 1d})--CCSDT(Q)/cc-pVTZ({no \em f 1d})) results in a minor lowering of the asymmetric stretch by $\sim$1 cm$^{-1}$, and a further improvement in the basis set to the cc-pVQZ({no \em g 1f}) basis set has a similar effect. Unfortunately, carrying out the fully iterative CCSDTQ/cc-pVTZ calculations is beyond the capabilities of our current computational resources, these calculations involving  5$\times$10$^{9}$ and 10.5$\times$10$^{9}$ amplitudes in $C_{2v}$ and $C_s$ symmetries, respectively. 

It is appropriate to note, at this point, that --- in spite of the difficulties the potential energy surface of ozone presents for W4 theory --- W4 performs very well for the thermochemistry of O$_3$. In particular, the post-CCSD(T) components are practically converged at the W4 level, as evident from very small differences between W4 and W4.3 theories. The post-CCSD(T) contributions to the total atomization energy of ozone in W4 and W4.3 theories are: $\hat{T}_3-$(T) $=$ -1.34 and -1.37; $\hat{T}_4=$ 3.81 and 3.84; and $\hat{T}_5=$ 0.41 and 0.43.\cite{O3} In addition, W4.3 includes a $\hat{T}_6$ contribution of 0.05 kcal/mol. This suggests that while basis set convergence is faster for second-order properties ({\em e.g.} frequencies) than for the energy, n-particle treatment convergence is arguably slower.

\subsection{Electrical properties\label{sec:elec}}
Fairly accurate experimental dipole moments are available for some of the molecules considered in the present work; however, for the higher-order properties ($\alpha$ and $\beta$) accurate experimental data are much more scarce. Furthermore, direct comparison between theory and experiment is hampered by the fact that the theoretical values correspond to the static limit, whereas the experimental values are generally frequency-dependent. Thus, we will not attempt to quantify the reliability of W4 theory against experimental data; our intent is rather to investigate the effects of core-valence and post-CCSD(T) contributions on the electrical properties. 

Table \ref{tab:dip} gives the theoretical and experimental dipole moments. Core-valence and post-CCSD(T) corrections have no noticeable effect on systems dominated by dynamical correlation effects, and relatively little effects on the strongly multireference systems. 
The dipole moments at the frozen-core, non-relativistic CCSD(T) basis set limit (W4[up to CCSD(T)$_{fc,nr}$] method) agree very well with the full W4 values (RMSD=0.005 a.u., and the largest deviation of 0.01 a.u. is seen for BeO). At the W4[up to CCSD(T)] level, the RMSD from W4 is 0.002 a.u., while the largest error of 0.005 a.u. is found for O$_3$. Adding the higher order triple excitations increases the errors for the nonhydrides: in particular, a dramatic increase is seen for BN (from -0.003 to 0.05 a.u.). We note that for BN, the $\hat{T}_3-$(T) excitations raise $\mu$ by 0.05 a.u. and the $\hat{T}_4$ excitations reduce it by 0.04 a.u. At the W4[up to CCSDTQ] level, the largest errors of -0.014 and +0.012 a.u. are seen for BeO and BN, respectively, and the RMSD from W4 is 0.06 a.u. 

Table \ref{tab:pol} shows the theoretical polarizabilities together with the available experimental data. The RMSDs from the full W4 values are 2.6, 2.9, 0.5, and 0.2 a.u. for the W4[up to CCSD(T)$_{fc,nr}$], W4[up to CCSD(T)], W4[up to CCSDT], and W4[up to CCSDTQ] methods, respectively. However, for most of the systems in Table \ref{tab:pol}, the polarizabilities at the W4[up to CCSD(T)$_{fc,nr}$] level agree to within 0.1 a.u. with the W4 values. The main exceptions are $\alpha_{zz}$ of BeO, and $\alpha_{zz}$ and $\alpha_{xx}$ of C$_2$, which are overestimated by as much as 1.5, 5.4, and 13.0 a.u., respectively. Upon excluding C$_2$ and BeO the RMSD of the W4[up to CCSD(T)$_{fc,nr}$] method drops to 0.13 a.u. The W4[up to CCSD(T)$_{fc,nr}$] and W4[up to CCSD(T)] methods show very similar performance. Inclusion of $\hat{T}_3-$(T) excitations dramatically reduces the said errors for C$_2$ to 0.1 and 1.0 a.u., respectively, and the RMSD for the entire set drops to 0.5 a.u. Nevertheless, at this level errors of -1.0 and 2.1 a.u. are still seen for $\alpha_{xx}$ and $\alpha_{zz}$ of BeO, respectively. These are reduced to -0.3 and 1.0 a.u., respectively, by inclusion of the $\hat{T}_4$ excitations in the W4[up to CCSDTQ] method. 

Table \ref{tab:hypol} lists the theoretical W4 (and related) hyperpolarizabilities. Similarly to the polarizabilities, for most systems, the first hyperpolarizabilities at the W4[up to CCSD(T)$_{fc,nr}$] level agree with the W4 values to within 0.1 a.u. The main exceptions are O$_3$ ($\beta_{yyz}$) and CS ($\beta_{zzz}$); excluding these, the W4[up to CCSD(T)$_{fc,nr}$] method attains an RMSD of 0.3 a.u. from the W4 values. Inclusion of the core-valence and relativistic contributions (in the W4[up to CCSD(T)] method) reduces the said RMSD to 0.1 a.u., and inclusion of post-CCSD(T) excitations offers little improvement. Turning our attention to the more problematic systems, we note that at the W4[up to CCSD(T)$_{fc,nr}$] level $\beta_{zzz}$ of CS is $\sim$3 a.u. removed from the W4 value and that core-valence and post-CCSD(T) corrections do not significantly affect this result. The W4[up to CCSD(T)$_{fc,nr}$] method gives the wrong sign for the $\beta_{yyz}$ component of O$_3$, {\em i.e.} it overestimates the W4 value by as much as $\sim$9 a.u. Inclusion of $\hat{T}_3-$(T) excitations reduces this error to $\sim$2 a.u. and further inclusion of $\hat{T}_4$ excitations reduces the error to -0.2 a.u.

\subsection{Conclusions}
For a chemically diverse set of 28 first- and second-row diatomic molecules for which very accurate experimental spectroscopic constants are available W4 theory is capable of spectroscopically accurate predictions. Specifically, the RMSD from experiment are 0.04 pm for $r_e$, 1.03 cm$^{-1}$ for $\omega_e$, 0.20 cm$^{-1}$ for $\omega_e x_e$, 0.10 cm$^{-1}$ for $\omega_e y_e$, and 0.001 cm$^{-1}$ for $\alpha_e$.

Core-valence and post-CCSD(T) contributions have little effect on $\omega_e x_e$ and  $\alpha_e$,  and especially on $\omega_e y_e$. As for $r_e$ and $\omega_e$, higher-order triple excitations, $\hat{T}_3-$(T), improve agreement with experiment for the monohydride systems, but their inclusion (in the absence of $\hat{T}_4$) generally worsens agreement with experiment for the nonhydride diatomics. For instance, upon inclusion of the $\hat{T}_3-$(T) corrections the RMSD in $\omega_e$ reduces from 5.1 to 2.8 cm$^{-1}$ for the subset of monohydride systems, and increases from 11.2 to 15.2 cm$^{-1}$ for the subset of nonhydride systems. 

Connected quadruple excitations, $\hat{T}_4$, are essential to obtain $r_e$ and $\omega_e$ with RMSD of $\sim$0.05 pm and $\sim$2 cm$^{-1}$, respectively. For example, inclusion of the $\hat{T}_4$ excitations reduces the RMSD in $\omega_e$ from 2.8 to 0.6 cm$^{-1}$ for the subset of monohydrides, and from 15.2 to 2.4 cm$^{-1}$ for the subset of nonhydrides. The RMSD in $r_e$ are reduced from 0.04 to 0.03 pm for the hydrides subset, and from 0.30 to 0.06 pm for the nonhydrides subset. Moreover, inclusion of $\hat{T}_4$ excitations results in systematic trends: namely $r_e$ universally increase (by up to 0.5 pm), $\omega_e$ universally decrease (by up to 31.5 cm$^{-1}$), and $\omega_ex_e$ universally increase (by up to 1.0 cm$^{-1}$). 

Connected quintuple excitations, $\hat{T}_5$, are of minor importance for $r_e$ and $\omega_e x_e$, but are spectroscopically significant for $\omega_e$. Inclusion of a CCSDTQ5/cc-pVDZ({no \em d}) correction term cuts the RMSD in the harmonic frequencies for our set of 28 diatomic molecules by 50\% from 2.0 to 1.0 cm$^{-1}$, practically all of the improvement being due to the nonhydride systems. 

DBOC corrections result in systematic trends and are of importance mainly for $r_e$ and $\omega_e$ of the hydride systems, something more pronounced for the lighter hydrides within each row. For example, for our set of monohydrides DBOC universally increase $r_e$ by 0.06--0.01 pm and universally decrease $\omega_e$ by 2.1--0.3 cm$^{-1}$. 

For the pathologically multireference systems (BeO and BN) obtaining near-spectroscopic accuracy requires: (i) augmenting the correlation-consistent basis sets on N and O in the valence post-CCSD(T) steps in W4 theory, and (ii) extrapolating the CCSD(T) inner-shell contribution in W4 theory from the aug-cc-pwCV\{Q,5\}Z basis set pair. 

The present study also considers water, carbon dioxide, and ozone. The experimental geometries and harmonic frequencies (with the exception of the asymmetric mode of ozone) are reproduced relatively accurately at the W4 level. However, the asymmetric stretch of ozone, which exhibits severe nondynamical character, is underestimated by as much as 20 cm$^{-1}$ at the W4 level. This deviation is reduced by more than 50\% when the connected quadruples, $\hat{T}_4$, term in W4 theory is calculated with the cc-pVTZ({no \em f 1d}) basis set. 

We also calculated the dipole moments ($\mu$), polarizabilities ($\alpha$), and first hyperpolarizabilities ($\beta$). We find that: (i) excluding the pathologically multireference systems, the post-CCSD(T) contributions are of little importance for these electrical properties; (ii) the core-valence and DBOC contributions are also of little importance in most cases; and (iii) that the sensitivity to post-CCSD(T) effects, particularly for systems with significant nondynamical correlation, increases with increasing order of derivative with respect to the static electric field, {\em i.e.}, in the order: $\mu<\alpha<\beta$. 

\section*{Acknowledgments}
Research at Weizmann was funded by the Israel Science Foundation (grant 709/05), the Helen and Martin Kimmel Center for Molecular Design, and the Weizmann Alternative Energy Research Initiative (AERI). JMLM is the incumbent of the Baroness Thatcher Professorial Chair of Chemistry and a member {\em ad personam} of the Lise Meitner-Minerva Center for Computational Quantum Chemistry. 

\section*{Supporting Information}
Diagnostics for the importance of nondynamical correlation can be found in Table S-I. Tables S-II through S-V give the $E^{{\rm core-valence}}$, $E^{\hat{T}_3-{\rm (T)}}$, $E^{\hat{T}_4}$, and $E^{\hat{T}_5}$ energy contributions as a function of $r$. Table S-VI gives theoretical and experimental second-order anharmonic corrections for the set of nonhydride diatomic molecules.

\clearpage
\squeezetable
\begin{table}
\caption{Equilibrium bond distances ($r_e$, pm) of diatomic molecules obtained from a Dunham analysis of potential energy curves computed at the W4 and related levels.\label{tab:Re}}
\resizebox{0.70\textwidth}{!}{%
\begin{tabular}{l|cccccccccc}
\hline\hline
 & W4[up to & W4 & W4 & W4 & W4 & W4.2 & W4.3 & $\Delta$DBOC$^a$ & Expt.& Ref. \\ 
 & CCSD(T) & [up to & [up to & [up to & & & & & & \\
 & $_{fc,nr}$] & CCSD(T)] & CCSDT] & CCSDTQ] & & & & & & \\
 \hline
BH     & 123.25  & 122.90 & 122.91 & 122.91 & 122.91 & 122.91 & 122.92 & 0.060 & 122.95      &\cite{JMLM91}\\
CH     & 111.96  & 111.75 & 111.77 & 111.77 & 111.78 &        &        & 0.041 & 111.777     &\cite{op117}\\
NH     & 103.68  & 103.55 & 103.58 & 103.60 & 103.60 &        &        & 0.026 & 103.655     &\cite{op117}\\
OH     & 97.01   & 96.92  & 96.93  & 96.96  & 96.96  &        &        & 0.015 &  96.966     &\cite{HH}\\
HF     & 91.73   & 91.67  & 91.66  & 91.69  & 91.69  & 91.69  & 91.69  & 0.005 &  91.6984    &\cite{op117}\\
AlH    & 164.99  & 164.48 & 164.53 & 164.54 & 164.54 &        &        & 0.053 & 164.5362    &\cite{Bern93}\\
SiH    & 152.18  & 151.81 & 151.85 & 151.86 & 151.86 &        &        & 0.039 & 151.966     &\cite{Mori86}\\
PH     & 142.35  & 142.06 & 142.11 & 142.13 & 142.13 &        &        & 0.028 & 142.14      &\cite{Hiro84}\\
SH     & 134.21  & 133.96 & 133.99 & 134.01 & 134.01 &        &        & 0.020 & 134.0614    &\cite{Bern95}\\
HCl    & 127.62  & 127.41 & 127.42 & 127.44 & 127.44 & 127.44 & 127.45 & 0.013 & 127.46149   &\cite{Wats73}\\
N$_2$  & 109.89  & 109.65 & 109.58 & 109.73 & 109.75 & 109.75 & 109.75 & 0.001 & 109.768$_5$ &\cite{HH}\\
O$_2$  & 120.61  & 120.42 & 120.33 & 120.66 & 120.70 &        &        & 0.000 & 120.752     &\cite{HH}\\
F$_2$  & 140.93  & 140.81 & 140.75 & 141.17 & 141.18 & 141.20 & 141.23 & 0.000 & 141.193     &\cite{HH}\\
Si$_2$ & 225.09  & 224.08 & 224.22 & 224.52 & 224.54 &        &        & 0.002 & 224.6       &\cite{HH}\\
P$_2$  & 189.76  & 188.95 & 188.77 & 189.23 & 189.27 & 189.30 &        & 0.001 & 189.34      &\cite{HH}\\
S$_2$  & 189.14  & 188.57 & 188.55 & 188.85 & 188.87 &        &        & 0.001 & 188.92      &\cite{HH}\\
Cl$_2$ & 198.87  & 198.47 & 198.55 & 198.70 & 198.70 & 198.71 &        & 0.001 & 198.7$_9$   &\cite{HH}\\
BF     & 126.66  & 126.21 & 126.21 & 126.24 & 126.23 &        &        & 0.001 & 126.25$_9$  &\cite{HH}\\
CO     & 113.02  & 112.75 & 112.72 & 112.80 & 112.80 & 112.80 & 112.81 & 0.001 & 112.82427   &\cite{Wats73}\\
BCl    & 172.19  & 171.41 & 171.45 & 171.49 & 171.48 &        &        & 0.000 & 171.5283    &\cite{Suen82}\\
CS     & 153.86  & 153.31 & 153.30 & 153.45 & 153.45 & 153.45 &        & 0.001 & 153.4941    &\cite{HH}\\
SiO    & 151.42  & 150.82 & 150.78 & 150.95 & 150.93 &        &        & 0.001 & 150.9739    &\cite{HH}\\
PN     & 149.35  & 148.81 & 148.63 & 149.00 & 149.05 &        &        & 0.001 & 149.0866    &\cite{HH}\\
SO     & 148.23  & 147.87 & 147.77 & 148.05 & 148.08 &        &        & 0.000 & 148.1087    &\cite{HH}\\
AlF    & 166.19  & 165.40 & 165.40 & 165.42 & 165.42 &        &        & 0.001 & 165.4369    &\cite{HH}\\
ClF    & 162.74  & 162.52 & 162.55 & 162.75 & 162.76 &        &        & 0.001 & 162.8313    &\cite{HH}\\
AlCl   & 213.87  & 212.91 & 212.94 & 212.96 & 212.96 &        &        & 0.001 & 213.0113    &\cite{HH}\\
SiS    & 193.49  & 192.67 & 192.68 & 192.88 & 192.85 &        &        & 0.001 & 192.9264    &\cite{Thad07}\\
\hline
\multicolumn{9}{c}{error statistics for hydrides (10 systems)$^b$}\\
\hline
MSD  & 0.21 & -0.04 & -0.01 & 0.00 & 0.00 & & & & \\
MAD  & 0.21 & 0.04  & 0.03  & 0.03 & 0.03 & & & & \\
RMSD & 0.25 & 0.06  & 0.04  & 0.03 & 0.03 & & & & \\
\hline
\multicolumn{9}{c}{error statistics for nonhydrides  (18 systems)$^b$}\\
\hline
MSD  & 0.30& -0.23 & -0.26 & -0.05 & -0.04 & & & & \\
MAD  & 0.36& 0.23  & 0.26  & 0.05  & 0.04 & & & & \\
RMSD & 0.43& 0.27  & 0.30  & 0.06  & 0.05 & & & & \\
\hline
\multicolumn{9}{c}{error statistics for everything (28 systems)$^b$}\\
\hline
MSD  & 0.27& -0.16 & -0.17 & -0.04 & -0.03 & & & & \\
MAD  & 0.30& 0.17  & 0.18  & 0.05  & 0.04  & & & & \\
RMSD & 0.33& 0.22  & 0.25  & 0.05  & 0.04  & & & & \\
\hline\hline
\end{tabular}}
\begin{flushleft}
$^a$DBOC correction at the HF/AVTZ level. 
$^b$Mean signed deviations (MSD), mean absolute deviation (MAD), and root mean squared deviation (RMSD) from experiment, note that the DBOC corrections (cfr. footnote $a$) are included in the error statistics.   
\end{flushleft}
\end{table}
\clearpage

\squeezetable
\begin{table}
\caption{Harmonic frequencies ($\omega_e$, cm$^{-1}$) of diatomic molecules obtained from a Dunham analysis of potential energy curves computed at the W4 and related levels.\label{tab:We}}
\resizebox{0.70\textwidth}{!}{%
\begin{tabular}{l|ccccccccc}
\hline\hline
 & W4[up to & W4 & W4 & W4 & W4 & W4.2 & W4.3 & $\Delta$DBOC$^a$ & Expt.$^b$ \\
 & CCSD(T) & [up to & [up to & [up to & & & & & \\
 & $_{fc,nr}$] & CCSD(T)] & CCSDT] & CCSDTQ] & & & & & \\
 \hline
BH     & 2360.24&2370.79 & 2369.13 & 2369.08 & 2369.08 & 2369.10 & 2368.69 & -2.08 & 2366.7296\\
CH     & 2858.08&2865.99 & 2863.57 & 2862.53 & 2862.51 &  &  & -1.83 & 2860.7508\\
NH     & 3286.50&3292.03 & 3287.33 & 3284.65 & 3284.62 &  &  & -1.43 & 3282.7200\\
OH     & 3742.91&3746.58 & 3744.10 & 3739.55 & 3739.48 &  &  & -0.90 & 3737.7610\\
HF     & 4141.64&4143.57 & 4143.71 & 4138.87 & 4138.77 & 4138.79 & 4138.51 & -0.31 & 4138.3850\\
AlH    & 1688.45&1685.17 & 1682.54 & 1682.14 & 1682.14 &  &  & -1.10 & 1682.3747\\
SiH    & 2048.29&2047.07 & 2043.63 & 2042.91 & 2042.90 &  &  & -1.04 & 2042.5229\\
PH     & 2370.75&2370.79 & 2365.80 & 2364.41 & 2364.41 &  &  & -0.91 & 2363.7740\\
SH     & 2700.78&2701.91 & 2698.64 & 2696.42 & 2696.40 &  &  & -0.73 & 2696.2475\\
HCl    & 2994.17&2995.65 & 2994.23 & 2991.50 & 2991.48 & 2991.60 & 2990.94 & -0.53 & 2990.9248\\
N$_2$  & 2362.34&2371.91 & 2382.27 & 2361.83 & 2358.54 & 2357.89 & 2358.21 & 0.04 & 2358.5700\\
O$_2$  & 1605.48&1609.22 & 1617.24 & 1585.76 & 1581.78 &  &  & 0.08 & 1580.1610\\
F$_2$  & 931.17 &932.34 & 936.12 & 919.44 & 918.96 & 917.82 & 917.44 & 0.02 & 916.9290\\
Si$_2$ & 516.41 &517.67 & 516.23 & 511.23 & 510.78 &  &  & 0.00 & 510.9800\\
P$_2$  & 787.79 &792.23 & 797.62 & 784.26 & 782.87 & 782.10 &  & 0.00 & 780.7700\\
S$_2$  & 734.42 &735.99 & 736.93 & 728.16 & 727.48 &  &  & 0.00 & 725.7102\\
Cl$_2$ & 564.79 &565.65 & 564.48 & 561.28 & 561.24 & 560.94 &  & 0.00 & 559.7510\\
BF     & 1394.96&1405.01 & 1404.45 & 1403.02 & 1403.25 &  &  & 0.04 & 1402.1587\\
CO     & 2166.60&2176.25 & 2178.86 & 2170.22 & 2170.37 & 2170.16 & 2170.35 & 0.05 & 2169.7559\\
BCl    & 836.91 &843.27 & 841.90 & 840.74 & 840.93 &  &  & 0.03 & 840.2947\\
CS     & 1285.98&1292.50 & 1293.26 & 1284.95 & 1285.24 & 1285.45 &  & 0.02 & 1285.1546\\
SiO    & 1241.66&1248.53 & 1250.98 & 1241.56 & 1242.79 &  &  & 0.01 & 1241.5439\\
PN     & 1344.72&1352.54 & 1363.39 & 1341.62 & 1338.13 &  &  & 0.02 & 1336.9480\\
SO     & 1160.68&1163.56 & 1168.78 & 1152.64 & 1151.10 &  &  & 0.00 & 1150.7913\\
AlF    & 799.79 &803.09 & 802.84 & 802.36 & 802.44 &  &  & 0.00 & 802.3245\\
ClF    & 791.38 &792.11 & 791.38 & 784.96 & 784.81 &  &  & 0.00 & 783.4534\\
AlCl   & 481.75 &482.80 & 482.42 & 482.11 & 482.12 &  &  & 0.00 & 481.7747\\
SiS    & 752.61 &755.69 & 755.59 & 750.30 & 751.09 &  &  & 0.00 & 749.6456\\
\hline
\multicolumn{9}{c}{error statistics for hydrides (10 systems)$^c$}\\
\hline
MSD & 1.88& 4.66 & 1.96 & -0.10 & -0.13 & & & \\ 
MAD & 4.49& 4.66 & 2.15 & 0.48 & 0.46 & & & \\ 
RMSD& 4.81& 5.09 & 2.80 & 0.61 & 0.60 & & & \\ 
\hline
\multicolumn{9}{c}{error statistics for nonhydrides  (18 systems)$^c$}\\
\hline
MSD & 4.61& 9.11 & 11.57 & 1.67 & 0.97 & & & \\
MAD & 6.41& 9.11 & 11.57 & 1.69 & 1.00 & & & \\
RMSD& 8.67& 11.22 & 15.17 & 2.35 & 1.21 & & & \\
\hline
\multicolumn{9}{c}{error statistics for everything (28 systems)$^c$}\\
\hline
MSD  & 3.64& 7.52 & 8.14 & 1.04 & 0.58 & & & \\
MAD  & 5.73& 7.52 & 8.21 & 1.26 & 0.80 & & & \\
RMSD & 7.79& 9.78 & 12.69 & 1.99 & 1.03 & & & \\
\hline\hline
\end{tabular}}
\begin{flushleft}
$^a$DBOC correction at the HF/AVTZ level. 
$^b$From Ref.\cite{Irik07}.
$^c$See footnote $b$ of Table \ref{tab:Re}.
\end{flushleft}
\end{table}
\clearpage

\squeezetable
\begin{table}
\caption{First-order anharmonic corrections ($\omega_e x_e$, cm$^{-1}$) of diatomic molecules obtained from a Dunham analysis of potential energy curves computed at the W4 and related levels.\label{tab:WeXe}}
\resizebox{0.70\textwidth}{!}{%
\begin{tabular}{l|ccccccccc}
\hline\hline
 & W4[up to& W4 & W4 & W4 & W4 & W4.2 & W4.3 & $\Delta$DBOC$^a$ & Expt.$^b$ \\
 & CCSD(T) & [up to & [up to & [up to & & & & & \\
 & $_{fc,nr}$] & CCSD(T)] & CCSDT] & CCSDTQ] & & & & & \\
 \hline
BH    & 48.92& 49.16 & 49.33 & 49.33 & 49.33 & 49.34 & 49.33 & -0.02 & 49.33983\\
CH    & 63.66& 63.89 & 64.19 & 64.26 & 64.26 &     &     & -0.01 & 64.4387\\
NH    & 77.93& 78.07 & 78.53 & 78.74 & 78.74 &     &     & 0.01 & 79.04\\
OH    & 84.38& 84.48 & 84.72 & 85.01 & 85.02 &     &     & 0.00 & 84.8813\\
HF    & 89.73& 89.83 & 89.87 & 90.09 & 90.10 & 90.10 & 90.13 & 0.01 & 89.9432\\
AlH   & 28.75& 28.45 & 28.59 & 28.61 & 28.61 &     &     & -0.01 & 29.05098\\
SiH   & 35.67& 35.59 & 35.84 & 35.89 & 35.89 &     &     & -0.01 & 36.0552\\
PH    & 43.10& 43.09 & 43.46 & 43.58 & 43.58 &     &     & 0.00 & 43.907\\
SH    & 47.98& 48.01 & 48.25 & 48.40 & 48.40 &     &     & 0.00 & 48.742\\
HCl   & 51.89& 51.92 & 52.02 & 52.18 & 52.18 & 52.18 & 52.16 & 0.00 & 52.8\\
N$_2$ & 13.86& 13.90 & 13.64 & 14.14 & 14.24 & 14.26 & 14.24 & 0.00 & 14.324\\
O$_2$ & 11.03& 11.07 & 10.85 & 11.81 & 11.96 &     &     & 0.00 & 11.95127\\
F$_2$ & 11.45& 11.41 & 11.07 & 11.40 & 11.40 & 11.45 & 11.44 & 0.00 & 11.3221\\
Si$_2$& 1.93 & 1.94 & 1.92 & 2.02 & 2.03 &     &     & 0.00 & 2.02\\
P$_2$ & 2.72 & 2.71 & 2.68 & 2.83 & 2.84 & 2.86 &     & 0.00 & 2.835\\
S$_2$ & 2.73 & 2.73 & 2.70 & 2.85 & 2.86 &     &     & 0.00 & 2.8582\\
Cl$_2$& 2.61 & 2.59 & 2.62 & 2.67 & 2.67 & 2.68 &     & 0.00 & 2.69427\\
BF    & 11.71& 11.81 & 11.85 & 11.86 & 11.86 &     &     & 0.00 & 11.82106\\
CO    & 13.11& 13.17 & 13.12 & 13.33 & 13.30 & 13.30 & 13.27 & 0.00 & 13.28803\\
BCl   & 5.37 & 5.42 & 5.44 & 5.47 & 5.47 &     &     & 0.00 & 5.4917\\
CS    & 6.32 & 6.37 & 6.36 & 6.54 & 6.52 & 6.51 &     & 0.00 & 6.502605\\
SiO   & 5.84 & 5.90 & 5.83 & 6.02 & 5.99 &     &     & 0.00 & 5.97437\\
PN    & 6.56 & 6.57 & 6.44 & 6.76 & 6.81 &     &     & 0.00 & 6.8958\\
SO    & 6.16 & 6.17 & 6.09 & 6.47 & 6.51 &     &     & 0.00 & 6.4096\\
AlF   & 4.78 & 4.86 & 4.86 & 4.86 & 4.86 &     &     & 0.00 & 4.849915\\
ClF   & 4.86 & 4.85 & 4.84 & 4.95 & 4.95 &     &     & 0.00 & 4.9487\\
AlCl  & 2.08 & 2.10 & 2.10 & 2.11 & 2.11 &     &     & 0.00 & 2.101811\\
SiS   & 2.52 & 2.53 & 2.52 & 2.61 & 2.59 &     &     & 0.00 & 2.58623\\
\hline
\multicolumn{9}{c}{error statistics for hydrides (10 systems)$^c$}\\
\hline
MSD  & -0.62& -0.57 & -0.34 & -0.21 & -0.21 & & & \\
MAD  & 0.62& 0.57 & 0.34 & 0.27 & 0.27 & & & \\
RMSD & 0.68& 0.64 & 0.41 & 0.32 & 0.32 & & & \\
\hline
\multicolumn{9}{c}{error statistics for nonhydrides  (18 systems)$^c$}\\
\hline
MSD  & -0.18& -0.15 & -0.22 & -0.01 & 0.01 & & & \\
MAD  & 0.19& 0.17 & 0.22 & 0.05 & 0.03 & & & \\
RMSD & 0.28& 0.26 & 0.35 & 0.07 & 0.04 & & & \\
\hline
\multicolumn{9}{c}{error statistics for everything (28 systems)$^c$}\\
\hline
MSD  & -0.34& -0.30 & -0.26 & -0.08 & -0.07 & & & \\
MAD  & 0.35& 0.31 & 0.27 & 0.13 & 0.12 & & & \\
RMSD & 0.48& 0.45 & 0.39 & 0.21 & 0.20 & & & \\
\hline\hline
\end{tabular}}
\begin{flushleft}
$^a$DBOC correction at the HF/AVTZ level. 
$^b$From Ref.\cite{Irik07}.
$^c$See footnote $b$ of Table \ref{tab:Re}.
\end{flushleft}
\end{table}
\clearpage

\squeezetable
\begin{table}
\caption{Second-order anharmonic corrections ($\omega_e y_e$, cm$^{-1}$) of diatomic molecules obtained from a Dunham analysis of potential energy curves computed at the W4 and related levels.\label{tab:WeYe}}
\resizebox{0.70\textwidth}{!}{%
\begin{tabular}{l|ccccccccc}
\hline\hline
 & W4[up to & W4 & W4 & W4 & W4 & W4.2 & W4.3 & $\Delta$DBOC$^a$ & Expt.$^b$ \\
 & CCSD(T) & [up to & [up to & [up to & & & & & \\
 & $_{fc,nr}$] & CCSD(T)] & CCSDT] & CCSDTQ] & & & & & \\
\hline
BH     & 0.384  &0.369 & 0.359 & 0.360 & 0.360 & 0.359 & 0.367 & 0.000 & 0.362\\
CH     & 0.322  &0.314 & 0.301 & 0.298 & 0.300 &     &     & -0.002 & 0.3634\\
NH     & 0.290  &0.138 & 0.113 & 0.096 & 0.096 &     &     & 0.001 & 0.367\\
OH     & 0.654  &0.518 & 0.515 & 0.507 & 0.507 &     &     & 0.003 & 0.5409\\
HF     & 1.064  &1.052 & 1.063 & 1.056 & 1.056 & 1.056 & 1.052 & 0.002 & 0.92449\\
AlH    & 0.238  &-0.030 & -0.031 & -0.032 & -0.032 &     &     & -0.001 & 0.24762\\
SiH    & 0.149  &0.104 & 0.095 & 0.092 & 0.092 &     &     & -0.002 & 0.1254\\
PH     & 0.029  &0.012 & -0.007 & -0.015 & -0.014 &     &     & 0.001 & 0.1059\\
SH     & 0.065  &0.061 & 0.050 & 0.040 & 0.040 &     &     & 0.000 & 0.1124\\
HCl    & 0.021  &-0.018 & -0.020 & -0.027 & -0.027 & -0.028 & -0.028 & 0.001 & 0.21803\\
\hline
\multicolumn{9}{c}{error statistics for hydrides (10 systems)$^c$}\\
\hline
MSD & -0.015& -0.084 & -0.093 & -0.099 & -0.099&  &  & \\
MAD & 0.075&  0.112 & 0.121  & 0.126  & 0.125 &  &  & \\
RMSD& 0.094&  0.147 & 0.155  & 0.160  & 0.160 &  &  & \\
\hline
\multicolumn{9}{c}{error statistics for everything (28 systems)$^c$}\\
\hline
MSD & -0.005& -0.032 & -0.034 & -0.038 & -0.038&  &  & \\
MAD & 0.030&  0.045 & 0.049  & 0.051  & 0.051 &  &  & \\
RMSD& 0.061&  0.095 & 0.100  & 0.103  & 0.103 &  &  & \\
\hline\hline
\end{tabular}}
\begin{flushleft}
$^a$DBOC correction at the HF/AVTZ level. 
$^b$From Ref.\cite{Irik07}.
$^c$See footnote $b$ of Table \ref{tab:Re}.
\end{flushleft}
\end{table}
\clearpage

\squeezetable
\begin{table}
\caption{Vibration-rotation coupling constants ($\alpha_e$, cm$^{-1}$) of diatomic molecules obtained from a Dunham analysis of potential energy curves computed at the W4 and related levels.\label{tab:ALPHAe}}
\resizebox{0.70\textwidth}{!}{%
\begin{tabular}{l|ccccccccc}
\hline\hline
 & W4[up to & W4 & W4 & W4 & W4 & W4.2 & W4.3 & $\Delta$DBOC$^a$ & Expt.$^b$ \\
 & CCSD(T) & [up to & [up to & [up to & & & & &\\
 & $_{fc,nr}$] & CCSD(T)] & CCSDT] & CCSDTQ] &  &  &  &  &\\
\hline
BH      & 0.41852 & 0.42106 & 0.42207 & 0.42205 & 0.42205 & 0.42207 & 0.42216 & -0.00027 & 0.421565\\
CH      & 0.53115 & 0.53345 & 0.53503 & 0.53543 & 0.53544 &         &         & -0.00026 & 0.53654\\
NH      & 0.64301 & 0.64513 & 0.64754 & 0.64869 & 0.64871 &         &         & -0.00018 & 0.65038\\
OH      & 0.71515 & 0.71697 & 0.71830 & 0.72005 & 0.72009 &         &         & -0.00016 & 0.7242\\
HF      & 0.78693 & 0.78854 & 0.78893 & 0.79043 & 0.79047 & 0.79048 & 0.79048 & -0.00002 & 0.7933704\\
AlH     & 0.18372 & 0.18561 & 0.18625 & 0.18636 & 0.18636 &         &         & -0.00010 & 0.1870527\\
SiH     & 0.21455 & 0.21641 & 0.21741 & 0.21761 & 0.21761 &         &         & -0.00006 & 0.21814\\
PH      & 0.24869 & 0.25042 & 0.25180 & 0.25223 & 0.25223 &         &         & -0.00005 & 0.25339\\
SH      & 0.27622 & 0.27774 & 0.27865 & 0.27922 & 0.27923 &         &         & -0.00005 & 0.2799\\
HCl     & 0.30344 & 0.30488 & 0.30532 & 0.30594 & 0.30595 & 0.30593 & 0.30588 & -0.00005 & 0.3069985\\
N$_2$   & 0.01697 & 0.01703 & 0.01682 & 0.01722 & 0.01729 & 0.01731 & 0.01729 &  0.00000 & 0.017318\\
O$_2$   & 0.01512 & 0.01518 & 0.01501 & 0.01578 & 0.01590 &         &         &  0.00000 & 0.0159305\\
F$_2$   & 0.01244 & 0.01244 & 0.01225 & 0.01259 & 0.01260 & 0.01264 & 0.01263 &  0.00000 & 0.0125952\\
Si$_2$  & 0.00128 & 0.00130 & 0.00130 & 0.00134 & 0.00134 &         &         &  0.00000 & 0.00135\\
P$_2$   & 0.00142 & 0.00143 & 0.00141 & 0.00146 & 0.00147 & 0.00148 &         &  0.00000 & 0.00149\\
S$_2$   & 0.00153 & 0.00155 & 0.00154 & 0.00159 & 0.00159 &         &         &  0.00000 & 0.00159754\\
Cl$_2$  & 0.00145 & 0.00145 & 0.00145 & 0.00147 & 0.00147 & 0.00148 &         &  0.00000 & 0.001516\\
BF      & 0.01888 & 0.01901 & 0.01905 & 0.01906 & 0.01906 &         &         &  0.00000 & 0.01904848\\
CO      & 0.01733 & 0.01741 & 0.01737 & 0.01754 & 0.01752 & 0.01752 & 0.01750 &  0.00000 & 0.01750513\\
BCl     & 0.00667 & 0.00674 & 0.00677 & 0.00679 & 0.00679 &         &         &  0.00000 & 0.0068124\\
CS      & 0.00580 & 0.00585 & 0.00585 & 0.00594 & 0.00593 & 0.00593 &         &  0.00000 & 0.00591835\\
SiO     & 0.00494 & 0.00499 & 0.00496 & 0.00505 & 0.00504 &         &         &  0.00000 & 0.00503784\\
PN      & 0.00534 & 0.00537 & 0.00528 & 0.00547 & 0.00550 &         &         &  0.00000 & 0.0055337\\
SO      & 0.00557 & 0.00560 & 0.00556 & 0.00575 & 0.00577 &         &         &  0.00000 & 0.0057508\\
AlF     & 0.00490 & 0.00499 & 0.00500 & 0.00500 & 0.00500 &         &         &  0.00000 & 0.004984261\\
ClF     & 0.00425 & 0.00426 & 0.00426 & 0.00433 & 0.00433 &         &         &  0.00000 & 0.0043385\\
AlCl    & 0.00159 & 0.00161 & 0.00161 & 0.00162 & 0.00162 &         &         &  0.00000 & 0.001611082\\
SiS     & 0.00143 & 0.00145 & 0.00145 & 0.00148 & 0.00147 &         &         &  0.00000 & 0.00147313\\
\hline  
\multicolumn{9}{c}{error statistics for hydrides (10 systems)$^c$}\\
\hline
MSD  &  -0.00514 & -0.00325 & -0.00214 & -0.00147 & -0.00146  & & & \\
MAD  &   0.00514 &  0.00325 &  0.00219 &  0.00152 &  0.00150  & & & \\
RMSD &   0.00549 &  0.00379 &  0.00279 &  0.00192 &  0.00190  & & & \\
\hline  
\multicolumn{9}{c}{error statistics for nonhydrides  (18 systems)$^c$}\\
\hline
MSD  &  -0.00016 & -0.00012 & -0.00016 & -0.00002 & -0.00001  & & & \\
MAD  &   0.00016 &  0.00012 &  0.00016 &  0.00003 &  0.00002  & & & \\
RMSD &   0.00024 &  0.00021 &  0.00028 &  0.00005 &  0.00002  & & & \\
\hline  
\multicolumn{9}{c}{error statistics for everything (28 systems)$^c$}\\
\hline
MSD  &  -0.00194 & -0.00124 & -0.00087 & -0.00054 & -0.00053  & & & \\
MAD  &   0.00194 &  0.00124 &  0.00089 &  0.00056 &  0.00055  & & & \\
RMSD &   0.00341 &  0.00236 &  0.00174 &  0.00119 &  0.00118  & & & \\
\hline\hline
\end{tabular}}
\begin{flushleft}
$^a$DBOC correction at the HF/AVTZ level. 
$^b$From Ref.\cite{Irik07}.
$^c$See footnote $b$ of Table \ref{tab:Re}.
\end{flushleft}
\end{table}
\clearpage

\squeezetable
\begin{table}
\caption{Spectroscopic constants ($r_e$ in pm, all the rest in cm$^{-1}$) for the 12-electron isoelectronic series C$_2$, BN, and BeO obtained from a Dunham analysis of potential energy curves computed at the W4 and related levels.$^a$\label{tab:12el}}
\begin{tabular}{l|ccccccccccc}
\hline\hline
 & W4[up to & W4 & W4 & W4 & W4 & W4$^b$ &W4$^{bc}$ & W4.2$^{bc}$ & W4.3$^{c}$ & Expt.$^d$ \\
 & CCSD(T) & [up to & [up to & [up to & & & & & &\\
 & $_{fc,nr}$] & CCSD(T)] & CCSDT]& CCSDTQ] & & & & & &\\
\hline
&\multicolumn{10}{c}{C$_2$ ($X~^1\Sigma^+_g$)}\\
\hline
$r_e$            & 124.40  & 124.04  & 124.00  & 124.19  & 124.21  &  & 124.22  & 124.22  & 124.22  & 124.2440 \\
$\omega_e$       & 1861.29 & 1872.11 & 1874.62 & 1859.09 & 1857.63 &  & 1857.43 & 1856.55 & 1855.84 & 1855.0142 \\
$\omega_e x_e$   & 12.51   & 12.48   & 12.63   & 13.44   & 13.45   &  & 13.45   & 13.50   & 13.53   & 13.5547 \\
$\omega_e y_e$   & -0.020  & -0.019  & -0.039  & -0.081  & -0.082  &  & -0.082  & -0.083  & -0.088  & -0.1321 \\    
$\alpha_e$       & 0.01724 & 0.01726 & 0.01728 & 0.01783 & 0.01783 &  & 0.01783 & 0.01788 & 0.01789 & 0.018013 \\     
\hline
&\multicolumn{10}{c}{BN ($a~^1\Sigma^+$)}\\
\hline
$r_e$            & 126.85  & 126.43  & 127.17  & 127.33  & 127.38  & 127.42  & 127.43  & 127.43 & 127.33 & 127.45081 \\
$\omega_e$       & 1742.85 & 1755.27 & 1719.62 & 1713.06 & 1711.08 & 1707.50 & 1707.28 & 1707.64& 1709.81 & 1705.4032 \\
$\omega_e x_e$   & 16.33   & 16.08   & 12.29   & 11.99   & 11.76   & 11.40   & 11.40   & 11.37  & 11.68 & 10.55338 \\
$\omega_e y_e$   & -0.505  & -0.511  & -0.004  & -0.013  & -0.015  & -0.015  & -0.015  & -0.024 & 0.036 & \\
$\alpha_e$       & 0.01669 & 0.01663 & 0.01665 & 0.01638 & 0.01627 & 0.01618 & 0.01617 & 0.01611 & 0.01632 & 0.013857 \\
\hline
&\multicolumn{10}{c}{BeO ($X~^1\Sigma^+$)}\\
\hline
$r_e$            & 133.63  & 132.89  & 132.85  & 132.98  & 132.88  & 133.01  & 133.07  & 133.07 &  & 133.09 \\
$\omega_e$       & 1479.86 & 1497.59 & 1507.02 & 1494.41 & 1501.65 & 1493.43 & 1488.55 & 1488.55&  & 1487.32 \\
$\omega_e x_e$   & 12.05   & 12.14   & 10.95   & 12.12   & 12.34   & 11.78   & 11.69   & 11.66  &  & 11.83 \\
$\alpha_e$       & 0.01889 & 0.01900 & 0.01810 & 0.01893 & 0.01889 & 0.01881 & 0.01884 & 0.01882&  & 0.0190 \\
\hline\hline
\end{tabular}
\begin{flushleft}
$^a$For C$_2$, BN, and BeO DBOC contributions are found to be insignificant and are not included in the theoretical values.
$^b$Diffuse functions are added to electronegative atoms (N and O) in all the valence post-CCSD(T) steps, see text.
$^c$The inner-shell correlation contribution is extrapolated from RCCSD(T)/aug-cc-pwCVQZ and RCCSD(T)/aug-cc-pwCV5Z calculations, see text.
$^d$C$_2$ from Ref.\cite{C2}, BN from Ref.\cite{RnB} and BeO from Ref.\cite{HH}.
\end{flushleft}
\end{table}
\clearpage

\squeezetable
\begin{table}
\caption{Theoretical$^a$ and experimental equilibrium geometries, harmonic frequencies, and fundamental frequencies (in pm, degrees, and cm$^{-1}$) for H$_2$O, CO$_2$, and O$_3$.\label{tab:tri}}
\begin{tabular}{l|ccccccccc}
\hline\hline
 & $r_e$ & $\theta_e$ & $\omega_1$ & $\omega_2$ & $\omega_3$ & $v_1$ & $v_2$ & $v_3$ & Ref.\\
\hline
&\multicolumn{9}{c}{CO$_2$}\\
\hline
W4    & 115.98 & & 1351.9 & 670.5 & 2396.4 & 1331.5 & 666.4 & 2350.2 & \\
Expt. & 116.00 & & 1353.8 & 672.9 & 2396.5 & 1332.9 & 672.9 & 2349.1 & \cite{Tef92}\\
\hline
&\multicolumn{9}{c}{H$_2$O}\\
\hline
W4         & 95.76 & 104.50 & 3834.6 & 1648.3 & 3945.7  & 3655.8 & 1595.3 & 3754.3 & \\  
W4$^b$     &       &        & 3835.2 & 1648.0 & 3946.3  & 3656.4 & 1595.2 & 3755.0 & \\
Expt.      & 95.72 & 104.52 & 3832.2 & 1648.5 & 3942.5  &        &        &        & \cite{Ply56}\\
Expt.      & 95.84 & 104.44 & 3831.8 & 1647.8 & 3942.4  &        &        &        & \cite{Jen88}\\
Expt.      &       &        &        &        &         & 3657.1 & 1594.7 & 3755.9 & \cite{Ten08}\\
\hline
&\multicolumn{9}{c}{O$_3$}\\
\hline
W4[up to CCSDT] &  &  & 1153.6 & 717.9 & 1085.7 & 1126.5 & 704.9 & 1042.1 & \\
W4[up to CCSDTQ] &  &  & 1137.0 & 715.5 & 1070.2 & 1105.3 & 701.5 & 1020.3 & \\
W4 & 127.15 & 116.82 & 1133.9 & 715.2 & 1067.1 & 1100.4 & 701.5 & 1016.0 &  \\
W4$^{*c}$ & & &  &  & 1078.1 & &  &  &  \\
Expt. & 127.17 & 116.78 & 1134.9 & 716.0 & 1089.2 & & & & \cite{Bar74}\\
Expt. & 127.17 & 116.75 & 1133.3 & 715.0 & 1087.3 & & & & \cite{Bar02}\\
\hline\hline
\end{tabular}
\begin{flushleft}
$^a$All the theoretical frequencies are computed at the W4-optimized geometries; unless otherwise indicated the DBOC contributions are not included in the theoretical values, cfr. footnote $b$.
$^b$Including a DBOC correction at the HF/aug'-cc-pVTZ level of theory.
$^c$W4$^{*}$=W4[up to CCSDT]+$\hat{T}_4$/cc-pVTZ({no \em f 1d})+$\hat{T}_5$/cc-pVDZ({no \em d}) see text.
\end{flushleft}
\end{table}
\clearpage

\squeezetable
\begin{table}
\caption{Dipole moments (in a.u., 1 a.u. = 2.5416 Debye).\label{tab:dip}}
\begin{tabular}{l|ccccccc}
\hline\hline
 &W4[up to & W4 & W4 & W4 & W4  & Expt. & Ref. \\
 & CCSD(T)& [up to & [up to & [up to & & & \\
 & $_{fc,nr}$]& CCSD(T)] & CCSDT] & CCSDTQ] & & &\\
\hline
BH      & 0.550& 0.552 & 0.551 & 0.551 &       &   0.50 & \cite{HH} \\
HF      & 0.709& 0.709 & 0.709 & 0.709 & 0.709 &   0.7185$^a$ & \cite{HH}\\
HCl     & 0.433& 0.430 & 0.429 & 0.429 & 0.429 &   0.4362$\pm$0.0001 & \cite{CRC85}\\
CO      & 0.044&  0.046 & 0.044 & 0.048 & 0.049 & 0.0432 & \cite{HH}\\
CS      & 0.774& 0.777  & 0.771 & 0.774 & 0.778 & 0.770 & \cite{HH}\\
BeO     & 2.496& 2.484  & 2.469 & 2.470 & 2.484   & 2.2680 & \cite{HH}\\
BN      & 0.796& 0.796  & 0.850 & 0.811 & 0.799 &  & \\
H$_2$O  & 0.729& 0.730  & 0.730 & 0.729 & 0.729 &  0.730$\pm$0.002 & \cite{CRC85}\\
O$_3$   & 0.218& 0.218  & 0.222 & 0.214 & 0.213 &  0.2100 & \cite{CRC85}\\
\hline
MSD$^b$  & 0.035 & 0.034 & 0.031 & 0.031 & 0.033 & &\\
MAD$^b$  & 0.038 & 0.038 & 0.035 &  0.035 & 0.038 & &\\
RMSD$^b$ & 0.083 & 0.079 & 0.074 & 0.074 & 0.079 & &\\
MSD$^c$  & 0.001 & 0.000 & 0.004 & -0.001 & & &\\
MAD$^c$  & 0.004 & 0.002 & 0.010 &  0.004 & & &\\
RMSD$^c$ & 0.005 & 0.002 & 0.018 &  0.006 & & &\\
\hline\hline
\end{tabular}
\begin{flushleft}
$^a$Ref.\cite{Mue70} gives 0.707 a.u.
$^b$Error statistics with respect to experiment.
$^c$Error statistics with respect to the W4 values.
\end{flushleft}
\end{table}
\clearpage

\squeezetable
\begin{table}
\caption{Static Polarizabilities (in a.u., 1 a.u. = 1.48185$\times$ 10$^{-25}$ cm$^3$)$^a$.\label{tab:pol}}
\resizebox{0.50\textwidth}{!}{%
\begin{tabular}{ll|ccccccc}
\hline\hline
 & & W4[up to & W4 & W4 & W4 & W4 & Expt. & Ref. \\
 & & CCSD(T) & [up to   & [up to & [up to & & & \\
 & & $_{fc,nr}$]& CCSD(T)] & CCSDT] & CCSDTQ] & & & \\
\hline
BH    & $\alpha_{xx}$  & 20.87& 20.78 & 20.78 & 20.77 & & & \\
      & $\alpha_{zz}$  & 23.14& 23.03 & 23.05 & 23.06 & & & \\
      & $\bar{\alpha}$ & 21.63& 21.53 & 21.53 & 21.53 & & & \\
HF    & $\alpha_{xx}$  & 5.15& 5.14  & 5.14  & 5.14  & 5.14  & 5.08 & $^b$\\
      & $\alpha_{zz}$  & 6.29& 6.28  & 6.28  & 6.29  & 6.29  & 6.40 & $^b$\\
      & $\bar{\alpha}$ & 5.53& 5.52  & 5.52  & 5.53  & 5.53  &  5.52$^c$  & $^b$ \\
HCl   & $\alpha_{xx}$  &16.68& 16.65 & 16.65 & 16.65 & 16.65 & & \\
      & $\alpha_{zz}$  &18.32& 18.30 & 18.30 & 18.30 & 18.30 & & \\
      & $\bar{\alpha}$ &17.23& 17.20 & 17.20 & 17.20 & 17.20 &  17.39$^d$ & \cite{Spa91}\\
CO    & $\alpha_{xx}$  &11.85& 11.81 & 11.80 & 11.81 & 11.81 &  11.86 & \cite{Sunil88} \\
      & $\alpha_{zz}$  &15.42& 15.38 & 15.36 & 15.40 & 15.42 &  15.51 & \cite{Sunil88} \\
      & $\bar{\alpha}$ &13.04& 13.00 & 12.99 & 13.01 & 13.01 &  13.08 & \cite{Sunil88} \\
CS    & $\alpha_{xx}$  &23.74& 23.62 & 23.63 & 23.62 & 23.63 &  &  \\
      & $\alpha_{zz}$  &37.93& 37.81 & 37.79 & 37.87 & 37.85 &  &  \\
      & $\bar{\alpha}$ &28.47& 28.35 & 28.35 & 28.37 & 28.37 &  &  \\
P$_2$ & $\alpha_{xx}$  &39.91& 39.66 & 39.66 & 39.65 & 39.65 & & \\
      & $\alpha_{zz}$  &67.86& 67.69 & 67.77 & 67.52 & 67.68 & & \\
      & $\bar{\alpha}$ &49.23& 49.00 & 49.03 & 48.94 & 49.00 & & \\
Cl$_2$& $\alpha_{xx}$  &24.86& 24.81 & 24.81 & 24.81 & 24.81 &  24.43  & \cite{LB51} \\
      & $\alpha_{zz}$  &41.74& 41.66 & 41.68 & 41.65 & 41.65 &  44.54  & \cite{LB51} \\
      & $\bar{\alpha}$ &30.48& 30.43 & 30.44 & 30.43 & 30.43 &  31.11$^e$  & \cite{LB51} \\
H$_2$O& $\alpha_{xx}$  & 9.20& 9.18  & 9.18  & 9.18  & 9.18  & & \\
      & $\alpha_{yy}$  & 9.84& 9.83  & 9.83  & 9.84  & 9.84  & & \\
      & $\alpha_{zz}$  & 9.48& 9.47  & 9.47  & 9.48  & 9.48  & & \\
      & $\bar{\alpha}$ & 9.50& 9.49  & 9.49  & 9.50  & 9.50  &  9.642$^f$  & \cite{Spa91} \\
CO$_2$& $\alpha_{xx}$  & 12.74 & 12.74 & 12.73 & 12.75 & 12.75 & \\
      & $\alpha_{zz}$  & 26.55 & 26.55 & 26.51 & 26.57 & 26.61 & \\
      & $\bar{\alpha}$ & 17.34 & 17.34 & 17.32 & 17.36 & 17.37 &  17.51$^g$ & \cite{Spa91} \\
N$_2$ & $\alpha_{xx}$  & 10.19& 10.15 & 10.15 & 10.16 & 10.16 & & \\
      & $\alpha_{zz}$  &14.78& 14.73 & 14.76 & 14.71 & 14.72 & & \\
      & $\bar{\alpha}$ &11.72& 11.68 & 11.69 & 11.68 & 11.68 &  11.744$^h$& \cite{CRC85} \\
F$_2$ & $\alpha_{xx}$  & 6.38& 6.37  & 6.37  & 6.37  & 6.37  & & \\
      & $\alpha_{zz}$  &12.34& 12.32 & 12.36 & 12.26 & 12.26 & & \\
      & $\bar{\alpha}$ & 8.37& 8.35  & 8.36  & 8.33  & 8.33  & 8.38&\cite{You72} \\
C$_2$ & $\alpha_{xx}$  &36.88& 38.68 & 22.86 & 23.82 & 23.86 & & \\
      & $\alpha_{zz}$  &31.80& 31.81 & 26.36 & 26.33 & 26.42 & & \\
      & $\bar{\alpha}$ &35.18& 36.39 & 24.03 & 24.66 & 24.71 & & \\
BeO   & $\alpha_{xx}$  & 31.46& 31.36 & 30.63 & 31.30 & 31.59 & & \\
      & $\alpha_{zz}$  & 33.39& 33.05 & 34.01 & 32.91 & 31.90 & & \\  
      & $\bar{\alpha}$ & 32.10& 31.92 & 31.76 & 31.84 & 31.70 & & \\
O$_3$ & $\alpha_{xx}$  & 11.75 & 11.72 & 11.72 & 11.73 & 11.73 & & \\
      & $\alpha_{yy}$  & 31.26 & 31.27 & 31.03 & 30.81 & 30.74 & & \\
      & $\alpha_{zz}$  & 14.18 & 14.14 & 14.10 & 14.11 & 14.11 & & \\
      & $\bar{\alpha}$ & 19.06 & 19.04 & 18.95 & 18.88 & 18.86 & 21.7 & \cite{CRC85}\\
\hline
&MSD$^i$  & 0.72 & 0.72 & 0.01 & 0.01 & & &\\
&MAD$^i$  & 0.73 & 0.75 & 0.17 & 0.06 & & &\\
&RMSD$^i$ & 2.59 & 2.89 & 0.47 & 0.20 & & &\\
\hline\hline
\end{tabular}}
\begin{flushleft}
$^a$$\bar{\alpha}=\frac{1}{3}(\alpha_{xx}+\alpha_{yy}+\alpha_{zz})$.
$^b$Expt. value with zero-point vibrational corrections taken from Ref.\cite{Sad81}.
$^c$Ref.\cite{CRC85} gives: $\bar{\alpha}$=5.4, and Ref.\cite{Spa91} gives $\bar{\alpha}$=5.601 a.u.
$^d$Ref.\cite{Bridge66} gives: $\bar{\alpha}$=17.55 a.u., Ref.\cite{LB51} gives: $\alpha_{xx}$=16.13, $\alpha_{zz}$=21.12, and $\bar{\alpha}$=17.75 a.u.
$^e$Ref.\cite{Bridge66} gives: $\bar{\alpha}$=31.1 a.u.
$^f$Ref.\cite{CRC85} gives: $\bar{\alpha}$=9.8 a.u.
$^g$Ref.\cite{Bridge66} gives: $\bar{\alpha}$=17.7 a.u., Ref.\cite{LB51} gives: $\bar{\alpha}$=17.88 a.u., and Ref.\cite{CRC85} gives: $\bar{\alpha}$=19.64 a.u.
$^h$Ref.\cite{LB51} gives: $\alpha_{xx}$= 9.79, $\alpha_{zz}$=16.06, and $\bar{\alpha}$=11.88 a.u., and Ref.\cite{Bridge66} gives: $\bar{\alpha}$=11.92 a.u.
$^i$Error statistics with respect to the W4 values. 
\end{flushleft}
\end{table}
\clearpage

\squeezetable
\begin{table}
\caption{Static Hyperpolarizabilities (in a.u., 1 a.u. = 8.6392$\times$10$^{-33}$cm$^4$statvolt$^{-1}$)$^a$.\label{tab:hypol}}
\begin{tabular}{ll|ccccc}
\hline\hline
 & &W4[up to & W4 & W4 & W4 & W4 \\
 & & CCSD(T)& [up to & [up to & [up to & \\
 & & $_{fc,nr}$]& CCSD(T)] & CCSDT] & CCSDTQ] & \\
\hline
BH & $\beta_{zzz}$     & 21.81 & 22.39 & 22.05 & 22.16 &  \\
   & $\beta_{xxz}$     & 69.95 & 68.92 & 68.88 & 69.06 \\
   & $\bar{\beta}$     & 70.85 & 69.27 & 69.43 & 69.57 \\
HF & $\beta_{zzz}$     & 8.75 & 8.77 & 8.79 & 8.82 & 8.83 \\
   & $\beta_{xxz}$     & 1.03 & 1.03 & 1.04 & 1.05 & 1.05 \\
   & $\bar{\beta}$     & 6.49 & 6.50 & 6.52 & 6.55 & 6.55 \\
HCl & $\beta_{zzz}$    & 9.77 & 9.78 & 9.81 & 9.81 & 9.81 \\
   & $\beta_{xxz}$     & 0.29 & 0.33 & 0.32 & 0.31 & 0.31 \\
   & $\bar{\beta}$     & 6.21 & 6.26 & 6.28 & 6.26 & 6.26 \\
CO & $\beta_{zzz}$     & 28.63& 28.45& 28.32& 28.67& 28.5 \\
   & $\beta_{xxz}$     &7.66 & 7.60 & 7.56 & 7.64 & 7.67 \\
   & $\bar{\beta}$     &26.37 & 26.19 & 26.06 & 26.38 & 26.31 \\
CS & $\beta_{zzz}$     &15.95& 15.22 & 15.62 & 15.90 & 12.81 \\
   & $\beta_{xxz}$     & 4.79  & 4.48  & 4.38  & 4.61  & 4.72 \\
   & $\bar{\beta}$     & 15.31 & 14.50 & 14.63 & 15.07 & 13.35 \\
H$_2$O & $\beta_{zzz}$ &11.86 & 11.90 & 11.90 & 11.96 & 11.96 \\
   & $\beta_{xxz}$     & 4.62 & 4.62 & 4.63 & 4.66 & 4.66 \\
   & $\beta_{yyz}$     & 9.48 & 9.51 & 9.53 & 9.56 & 9.56 \\
   & $\bar{\beta}$     & 15.58 & 15.62 & 15.64 & 15.70 & 15.70 \\
O$_3$ & $\beta_{zzz}$  & -17.32 & -17.22 & -17.40 & -17.43 & -17.44 \\
   & $\beta_{xxz}$     & -3.61 & -3.59 & -3.61 & -3.62 & -3.63 \\
   & $\beta_{yyz}$     & 1.66 & 1.92 & -5.32 & -7.22 & -7.04 \\
   & $\bar{\beta}$     &-11.56 & -11.33 & -15.80 & -16.96 & -16.87 \\
\hline
&MSD$^b$  & 0.77 & 0.70 & 0.22 & 0.18 &  \\
&MAD$^b$  & 0.77 & 0.70 & 0.32 & 0.20 &  \\
&RMSD$^b$ & 2.26 & 2.26 & 0.81 & 0.75 &  \\
\hline\hline
\end{tabular}
\begin{flushleft}
$^a$$\bar{\beta}=\frac{3}{5}(\beta_{xxz}+\beta_{yyz}+\beta_{zzz})$.
$^b$Error statistics with respect to the W4 values. 
\end{flushleft}
\end{table}
\clearpage

\begin{thebibliography}{99}
\bibitem{O3}The W4 values are taken from Ref.\cite{op200}, and due to computational resources constraints the CCSDTQ$-$CCSDT(Q) difference in W4.3 theory is calculated with the $spd$ part of the cc-pVTZ basis set instead of the regular cc-pVTZ basis set.
\end{thebibliography}
\end{document}